\documentclass[12pt,a4paper]{article}
\usepackage{t1enc} \usepackage{times}
\usepackage{amssymb}
\usepackage{eurosym}
\usepackage{amsmath}
\usepackage[mathscr]{euscript}
\usepackage{graphicx}
\usepackage{float}
\usepackage{hyperref}
\usepackage[totalwidth=437pt, totalheight=664pt]{geometry}

\DeclareMathAlphabet\scr{U}{scr}{m}{n}
\SetMathAlphabet\scr{bold}{U}{scr}{b}{n}
 \DeclareFontFamily{U}{scr}{\skewchar\font'177}%
 \DeclareFontShape{U}{scr}{m}{n}{<-6>rsfs5<6-8>rsfs7<8->rsfs10}{}%
 \DeclareFontShape{U}{scr}{b}{n}{<-6>rsfs5<6-8>rsfs7<8->rsfs10}{}%

\newcommand{\zu}{]\!]}
\newcommand{\mal}{\stackrel{\mbox{\tiny$\bullet$}}{}}

\newcommand{\rr}{\mathbb R}  
\newcommand{\rp}{\mathbb R _+}

\newcommand{\nn}{\mathbb N}

\newtheorem{satz}{Theorem}[section]
\newtheorem{prop}[satz]{Proposition}

\newtheorem{lemma}[satz]{Lemma}
\newtheorem{cor}[satz]{Corollary}
\newtheorem{@definition}[satz]{Definition}
\newenvironment{defi}{\begin{@definition}\rm}{\end{@definition}}
\newtheorem{@bsp}[satz]{Example}
\newenvironment{bsp}{\begin{@bsp}\rm}{\end{@bsp}}
\newtheorem{@assumption}[satz]{Assumption}
\newenvironment{assumption}{\begin{@assumption}\rm}{\end{@assumption}}
\newtheorem{@convention}[satz]{Convention}

\newtheorem{@remark}[satz]{Remark}
\newenvironment{bem}{\begin{@remark}\rm}{\end{@remark}}
\newenvironment{bemm}{\noindent {\bf Remarks.}}{}

\newcommand{\be}{\begin{enumerate}}
\newcommand{\ee}{\end{enumerate}}
\newcommand{\beq}{\begin{equation}}
\newcommand{\eeq}{\end{equation}}
\newcommand{\bea}{\begin{eqnarray}}
\newcommand{\eea}{\end{eqnarray}}
\newcommand{\beaa}{\begin{eqnarray*}}
\newcommand{\eeaa}{\end{eqnarray*}}

\newcommand{\bpf}{\noindent {\sc Proof.}\ }
\newcommand{\ep}{\hfill $\square $}
\newcommand{\F}{\scr F}

\renewcommand{\H}{\scr H}

\renewcommand{\L}{\scr L}

\newcommand{\E}{\scr E}

\newcommand{\B}{\scr B}

\newcommand{\apl}{\scr A^+_\mathrm{loc}}

\newcommand{\til}{\widetilde}

\renewcommand{\emptyset}{\varnothing}

\renewcommand{\epsilon}{\varepsilon}
\renewcommand{\theta}{\vartheta}
\renewcommand{\rho}{\varrho}

\newcommand{\eur}{\mbox{\begin{scriptsize}\euro \end{scriptsize}}}

\renewenvironment{thebibliography}[1]{%
\begin{oldthebibliography}{#1}%
\setlength{\baselineskip}{.9em}
\linespread{.9}
\small
\setlength{\parskip}{0ex}%
\setlength{\itemsep}{.1em}%
}%
{%
\end{oldthebibliography}%
}

\begin{document}
\title{Asymptotic Power Utility-Based\\ Pricing and Hedging\footnote{We thank an anonymous referee and an associate editor for their careful reading of the manuscript.}}
\author{Jan Kallsen\footnote{Mathematisches Seminar,
Christian-Albrechts-Universit\"at zu Kiel,
Westring 383,
D--24118 Kiel, Germany,
(e-mail: kallsen@math.uni-kiel.de).}  
\quad Johannes Muhle-Karbe\footnote{Corresponding author. Departement Mathematik,
ETH Z\"urich,
R\"amistrasse 101,
CH--8092 Z\"urich, Switzerland, and Swiss Finance Institute.
(e-mail: johannes.muhle-karbe@math.ethz.ch). Partially supported by the National Centre of Competence in Research Financial Valuation and Risk Management (NCCR FINRISK), Project D1 (Mathematical Methods in Financial Risk Management), of the Swiss National Science Foundation (SNF).}\\
\quad Richard Vierthauer \footnote{Mathematisches Seminar,
Christian-Albrechts-Universit\"at zu Kiel,
Westring 383,
D--24118 Kiel, Germany,
(e-mail: vierthauer@math.uni-kiel.de).}  
}
\date{}
\maketitle

\begin{abstract}
Kramkov and S\^irbu \cite{kramkov.sirbu.06, kramkov.sirbu.06b} have shown that first-order approximations of power utility-based prices and hedging strategies for a small number of claims can be computed by solving a mean-variance hedging problem under a specific equivalent martingale measure and relative to a suitable numeraire. For power utilities, we propose an alternative representation that avoids the change of numeraire. More specifically, we characterize the relevant quantities using semimartingale characteristics similarly as in \v{C}ern\'y and Kallsen \cite{cerny.kallsen.05} for mean-variance hedging. These results are illustrated by applying them to exponential L\'evy processes and stochastic volatility models of Barndorff-Nielsen and Shephard type \cite{barndorff.shephard.01}. We find that asymptotic utility-based hedges are virtually independent of the investor's risk aversion. Moreover, the price adjustments compared to the Black-Scholes model turn out to be almost linear in the investor's risk aversion, and surprisingly small unless very high levels of risk aversion are considered.\\

\end{abstract}

\newpage

\section{Introduction}\setcounter{equation}{0}
In incomplete markets, derivative prices cannot generally be based on perfect
replication. A number of alternatives have been suggested in the literature, relying, e.g., 
on superreplication, mean-variance hedging, calibration of parametric
families, utility-based concepts, or ad-hoc approaches. This paper focuses on utility indifference prices as studied by Hodges and Neuberger \cite{hodges.neuberger.89} and many others. They make sense  for over-the-counter trades of a fixed quantity of contingent claims. Suppose that a client approaches a potential seller
in order to buy $q$ European-style contingent claims maturing at $T$. The seller is supposed
to be a utility maximizer with given preference structure. She will enter into the contract only if her
maximal expected utility is increased by  the trade. The utility indifference price
is the lowest acceptable premium for the seller. If the trade is made, the seller's optimal
position in the underlyings changes due to the presence of the option. This adjustment in the optimal
portfolio process is called utility-based hedging strategy for the claim. Both
the utility indifference price and the corresponding utility-based hedging strategy are 
typically hard to compute even if relatively simple incomplete market models are considered.
A reasonable way out for practical purposes is to consider approximations for small $q$,
i.e., the limiting structure for  small numbers of contingent claims.
Extending earlier work on the limiting price, Kramkov and S\^irbu \cite{kramkov.sirbu.06,kramkov.sirbu.06b} show that first-order approximations of the utility indifference price and the utility-based hedging strategy
can be expressed in terms of a Galtchouk-Kunita-Watanabe (GKW) decomposition of the
claim after changing both the numeraire and  the underlying probability measure.

From a slightly different perspective one may say that
Kramkov and S\^irbu \cite{kramkov.sirbu.06, kramkov.sirbu.06b}
relate utility indifference pricing and hedging asymptotically to some
mean-variance hedging
problem. In this representation, the $L^2$-distance between payoff and
terminal wealth of
approximating portfolios needs to be considered relative to both a new
numeraire and a new probability measure. 

This differs from related results for exponential utility (see \cite{mania.schweizer.05,becherer.06,kallsen.rheinlaender.08}), where no numeraire change is necessary. In the present study, we show that the numeraire change can also be avoided for power utilities, which constitute the most popular and tractable ones on the positive real line, i.e., in the setup of \cite{kramkov.sirbu.06, kramkov.sirbu.06b}. This allows to examine directly how the dynamics of the underlying change to account for utility-based rather than mean-variance hedging, and also allows to apply directly a number of explicit resp.\ numerical results from the literature. The key idea is to consider an equivalent mean-variance hedging problem relative to the original numeraire but under yet another probability measure. More specifically, the solution of \cite{kramkov.sirbu.06, kramkov.sirbu.06b} for a contingent claim $H$ corresponds to a quadratic hedging problem of the form
\beq\label{e:mvhedge1}
\min_{c,\varphi}E_{Q^\$}\!\left(\bigg({c+\varphi\mal S_T-H\over
N_T}\bigg)^2\right)
\eeq
with some numeraire process $N$ and some martingale measure $Q^\$ $.
If we define a new measure $P^{\eur}$ via
$${dP^{\eur}\over dQ^\$}:=\frac{1/N_T^2}{E_{Q^\$}(1/N_T^2)},$$
the mean-variance hedging problem \eqref{e:mvhedge1} can evidently be rewritten as
\beq\label{e:mvhedge2}
E_{Q^\$}(1/N_T^2)\min_{c,\varphi}E_{P^{\eur}}\!\left((c+\varphi\mal S_T-H)^2)\right),
\eeq
where we minimize again over some a set of initial endowments $c$ and trading strategies $\varphi$. Replacing \eqref{e:mvhedge1} by 
\eqref{e:mvhedge2} constitutes the key idea underlying our approach. For a related transition in the quadratic hedging literature compare
\cite{gourieroux.al.98} with and \cite{schweizer.94, rheinlaender.schweizer.97, cerny.kallsen.05} without numeraire change. Since the stock is not a martingale in the reformulation \eqref{e:mvhedge2}, the Galtchouk-Kunita-Watanabe decomposition does not lead to the solution. Instead, representations as in \cite{schweizer.94} or, more generally, \cite{cerny.kallsen.05} can be used to obtain concrete formulas, which are provided in Theorem 4.7 of this paper. On a rigorous mathematical level, we do not consider mean-variance hedging problems because the expression in Theorem 4.7 is the solution to such a hedging problem only under additional regularity which does not hold in general. Instead, we show in a more direct fashion that the solution of \cite{kramkov.sirbu.06, kramkov.sirbu.06b} can be expressed as in Theorem 4.7. 

In order to illustrate the applicability of our results and shed light on the role of the investor's risk aversion for power utility-based pricing and hedging, we consider exponential L\'evy processes and the stochastic volatility model of Barndorff-Nielsen and Shephard \cite{barndorff.shephard.01} as examples. For these processes, all technical assumptions can be verified directly in terms of the model parameters. Moreover, results for the related mean-variance hedging problem (cf.\ \cite{hubalek.al.05, cerny.05, kallsen.pauwels.09a, kallsen.vierthauer.09}) can be adapted to obtain first-order approximations to utility-based prices and hedging strategies explicitly up to some numerical integrations. Using parameters estimated from an equity index time series, we find that the asymptotic utility-based hedging strategies are virtually independent of the investor's risk aversion, which holds exactly for exponential investors. Moreover, the risk premia per option sold turn out to be almost linear in the investor's (absolute) risk aversion, which again holds exactly for exponential utilities. Hence, these examples suggest very similar pricing and hedging implications of both exponential and power utilities: risk aversion barely influences the optimal hedges, and enters linearly into the first-order risk-premia. Similarly as in \cite{henderson.hobson.02, henderson.02} in the context of basis risk, we find that the
 price adjustments are negligibly small for the levels of risk aversion typically considered in the literature. In particular, surprisingly high levels of risk aversion are needed to obtain bid- and ask prices below and above the Black-Scholes price, respectively.

The remainder of the paper is organized as follows. After briefly recalling the general theory of power utility-based pricing and hedging in Section 2, we review the asymptotic results of Kramkov and S\^irbu \cite{kramkov.sirbu.06,kramkov.sirbu.06b}. As a byproduct we derive a feedback formula for the utility-based hedging strategy. Subsequently, we develop our alternative representation in Section 4. Throughout, we explain how to apply the general theory to exponential L\'evy processes and the stochastic volatility model of Barndorff-Nielsen and Shephard \cite{barndorff.shephard.01}. A concrete numerical example is considered in Section 6. Finally, the appendix summarizes notions and results concerning semimartingale calculus for the convenience of the reader.

Unexplained notation is generally used as in the monograph of Jacod and Shiryaev \cite{js.03}. In particular, for a semimartingale $X$, we denote by $L(X)$ the predictable $X$-integrable processes and by $\varphi \mal X$ the stochastic integral of $\varphi \in L(X)$ with respect to $X$. We write $\E(X)$ for the stochastic exponential of a semimartingale $X$ and denote by $\L(Z):=\frac{1}{Z_{-}}\mal Z$ the \emph{stochastic logarithm} of a semimartingale $Z$ satisfying $Z, Z_{-} \neq 0$. For semimartingales $X$ and $Y$, $\langle X,Y \rangle$ represents the predictable compensator of $[X,Y]$, provided that the latter is a special semimartingale (cf.\ \cite[p.\ 37]{jacod.79}). Finally, we write $c^{-1}$ for the \emph{Moore-Penrose pseudoinverse} of a matrix or matrix-valued process $c$ (cf.\ \cite{albert.72}) and denote by $E_d$ the identity matrix on $\rr^d$.  

\section{Utility-based pricing and hedging}\setcounter{equation}{0}
Our mathematical framework for a frictionless market model is as follows. Fix a terminal time $T > 0$ and a filtered probability space $(\Omega,\scr{F},(\scr{F}_t)_{t \in [0,T]},P)$ in the sense of \cite[I.1.2]{js.03}. For ease of exposition, we assume that $\scr{F}_T=\scr{F}$ and $\scr{F}_0=\{\emptyset,\Omega\}$ up to null sets, i.e., all $\scr{F}_0$-measurable random variables are almost surely constant.

We consider a securities market which consists of $d+1$ assets, a bond and $d$ stocks. As is common in Mathematical Finance, we work in \textit{discounted} terms. This means we suppose that the bond has constant value $1$ and denote by $S=(S^1,\ldots,S^d)$ the \textit{discounted price process} of the $d$ stocks in terms of multiples of the bond. The process $S$ is assumed to be an $\rr^d$-valued semimartingale. 

\begin{bsp}
\begin{enumerate}
\item Throughout this article, we will illustrate our results by considering one-dimensional \emph{exponential L\'evy models}. This means that $d=1$ and $S=S_0\E(X)$ for a constant $S_0>0$ and a L\'evy process $X$ with L\'evy-Khintchine triplet $(b^X,c^X,F^X)$ relative to some truncation function $h$ on $\mathbb{R}$. We write
$$\psi^X(z)=zb^X+\frac{1}{2}z^2 c^X+\int (e^{zx}-1-zh(x))F^X(dx)$$ 
for the corresponding \emph{L\'evy exponent}, i.e., the function $\psi^X: i\mathbb{R^d} \to \mathbb{C}$ such that  $E(e^{zX_t})= \exp(t\psi^X(z))$. When considering exponential L\'evy models, we will always assume $S>0$, which is equivalent to $\Delta X>-1$ resp.\ the support of $F^X$ being concentrated on $(-1,\infty)$.
\item We will also consider the stochastic volatility model of Barndorff-Nielsen and Shephard \cite{barndorff.shephard.01} (henceforth \emph{BNS model}). Here $d=1$ and the return process $X$ driving $S=S_0\E(X)$ is modelled as
$$dX_t=\mu y_t dt+\sqrt{y_t}dW_t, \quad X_0=0,$$
for a constant $\mu \in \mathbb{R}$, a standard Brownian motion $W$, and an independent L\'evy-driven Ornstein-Uhlenbeck process $y$. The latter is given as the solution to the SDE
$$dy_t=-\lambda y_t dt + dZ_t, \quad y_0>0,$$
with some constant $\lambda>0$ and an increasing L\'evy process $Z$ with L\'evy-Khintchine triplet $(b^Z,0,F^Z)$ relative to a truncation function $h$ on $\mathbb{R}$.
\end{enumerate}
\end{bsp}

Self-financing \emph{trading strategies} are described by $\rr^d$-valued predictable stochastic processes $\varphi=(\varphi^1,\ldots,\varphi^d)$, where $\varphi^i_t$ denotes the number of shares of security $i$ held at time $t$. We consider an investor whose preferences are modelled by a \emph{power utility function} $u(x)=x^{1-p}/(1-p)$ with constant relative risk aversion $p \in \rp \backslash \{0,1\}$. Given an initial endowment $v>0$, the investor solves the \emph{pure investment problem}
\begin{equation}\label{e:pureinvestment}
U(v):=\sup_{\varphi \in \Theta(v)} E(u(v+\varphi \mal S_T)),
\end{equation}
where the set $\Theta(v)$ of \emph{admissible strategies} for initial endowment $v$ is given by
$$ \Theta(v) := \{\varphi \in L(S): v+\varphi \mal S \geq 0\}.$$

To ensure that the optimization problem \eqref{e:pureinvestment} is well-posed, we make the following two standard assumptions. 

\begin{assumption}\label{a:1}
There exists an \emph{equivalent local martingale measure}, i.e., a probability measure $Q \sim P$ such that $S$ is a local $Q$-martingale.
\end{assumption} 

\begin{assumption}\label{a:2} The maximal expected utility in the pure investment problem \eqref{e:pureinvestment} is finite, i.e., $U(v)<\infty$. \end{assumption}

\begin{bsp}\label{bsp:wellposed}
\begin{enumerate}
\item In a univariate exponential L\'evy model $S=S_0\E(X)>0$, Assumption \ref{a:1} is satisfied if $X$ is neither a.s.\ decreasing nor a.s.\ increasing. In this case, by \cite[Corollary 3.7]{nutz.10c}, Assumption \ref{a:2} holds if and only if $\int_{\{|x|>1\}} x^{1-p}F^X(dx)<\infty$, i.e., if and only if the return process $X$ has finite $(1-p)$-th moments.
\item By \cite[Theorem 3.3]{kallsen.muhlekarbe.10}, Assumptions \ref{a:1} and \ref{a:2} are always satisfied in the BNS model if the investor's risk aversion $p$ is bigger than $1$. For $p \in (0,1)$, they hold provided that
\begin{equation}\label{eq:bnsint}
\int_1^{\infty} \exp\left(\frac{1-p}{2p} \mu^2 \frac{1-e^{-\lambda T}}{\lambda}z\right)F^Z(dz)<\infty,
\end{equation}
i.e., if sufficiently large exponential moments of the driving L\'evy process $Z$ exist.
\end{enumerate} 
\end{bsp}

In view of \cite[Theorem 2.2]{kramkov.schachermayer.99}, Assumptions \ref{a:1} and \ref{a:2} imply that the supremum in \eqref{e:pureinvestment} is attained for some strategy $\widehat{\varphi} \in \Theta(v)$ with strictly positive wealth process $v+\widehat{\varphi} \mal S$. By Assumption \ref{a:1} and \cite[I.2.27]{js.03}, $v+\widehat{\varphi} \mal S_{-}$ is strictly positive as well and we can write
\begin{equation*}
v+\widehat{\varphi} \mal S = v \scr{E}(-\til{a}\mal S)
\end{equation*}
for the \emph{optimal number of shares per unit of wealth} 
$$-\til{a}:=\frac{\widehat{\varphi}}{v+\widehat{\varphi} \mal S_{-}},$$
which is independent of the initial endowment $v$ for power utility. Finally, \cite[Theorem 2.2]{kramkov.schachermayer.99} also establishes the existence of a \emph{dual minimizer}, i.e., a strictly positive supermartingale $\widehat{Y}$ with $\widehat{Y}_T=\scr{E}(-\til{a}\mal S)_T^{-p}$ such that $(v+\varphi\mal S)\widehat{Y}$ is a supermartingale for all $\varphi \in \Theta(v)$ and $(v+\widehat{\varphi} \mal S)\widehat{Y}$ is a true martingale. Alternatively, one can represent this object in terms of the \emph{opportunity process} $L:=L_0\E(K):=\E(-\til{a} \mal S)^p \widehat{Y}$ of the power utility maximization problem (cf.\ \cite{cerny.kallsen.05,kallsen.muhlekarbe.10} for motivation and more details).

The optimal strategy $\widehat{\varphi}$ as well as the joint characteristics of the assets and the opportunity process $L$ satisfy a semimartingale \emph{Bellman equation} (cf.\ \cite[Theorem 3.2]{nutz.09c}). In concrete models, this sometimes allows to determine $\widehat{\varphi}$ and $L$ by making an appropriate ansatz. 

\begin{bsp}\label{bsp:levyinvest}
\begin{enumerate}
\item Let $S=S_0 \E(X)>0$ for a non-monotone L\'evy process $X$ with finite $(1-p)$-th moments. Then it follows from \cite[Lemma 5.1]{nutz.10c} that there exists a unique maximizer $\widehat{\eta}$ of
$$\mathfrak{g}(\eta)=\eta b^X-\frac{p}{2}\eta^2 c^X+\int \left(\frac{(1+\eta x)^{1-p}-1}{1-p}-\eta h(x)\right)F^X(dx),$$
over the set $\scr{C}^0= \{\eta \in \mathbb{R}: F^X(x \in \mathbb{R}: \eta x <-1)=0\}$ of fractions of wealth invested into stocks that lead to nonnegative wealth processes. By \cite[Theorem 3.2]{nutz.10c}, the optimal number of shares per unit of wealth is given by
$$-\til{a}=\widehat{\eta}/S_{-},$$
with corresponding wealth process $v\E(-\til{a}\mal S)=v\E(\widehat{\eta} X)$ and opportunity process
$$L_t=\exp(a(T-t)), \quad \mbox{where } a=(1-p)\max_{\eta \in \scr{C}^0}\mathfrak{g}(\eta).$$
\item By \cite[Theorem 3.3]{kallsen.muhlekarbe.10}, it is also optimal to hold a constant fraction of wealth in stocks in the BNS model, namely $\widehat{\eta}=\mu/p$ (provided that the conditions of Example \ref{bsp:wellposed} are satisfied). The optimal number of shares per unit of wealth is then given by $-\til{a}=\widehat{\eta}/S_{-}$ with corresponding wealth process $v\E(\widehat{\eta} X)$, and opportunity process 
$$L_t=\exp(\alpha_0(t)+\alpha_1(t)v_t),$$
for
\begin{align*}
\alpha_1(t)=\frac{1-p}{2p}\mu^2 \frac{1-e^{-\lambda(T-t)}}{\lambda},\quad \alpha_0(t)=\int_t^T \psi^Z(\alpha_1(s))ds,
\end{align*}
where $\psi^Z$ denotes the L\'evy exponent of $Z$.
\end{enumerate}
\end{bsp}

In addition to the traded securities, we now also consider a non-traded European contingent claim with maturity $T$ and payment function $H$, which is an $\scr{F}_T$-measurable random variable. Following \cite{kramkov.sirbu.06,kramkov.sirbu.06b}, we assume that $H$ can be superhedged by some admissible strategy as, e.g., for European puts and calls.

\begin{assumption} \label{a:3}
$|H| \leq w+\varphi \mal S_T$ for some $w \in (0,\infty)$ and $\varphi \in \Theta(w)$.
\end{assumption}

If the investor sells $q$ units of $H$ at time $0$, her terminal wealth should be sufficiently large to cover the payment $-qH$ due at time $T$. This leads to the following definition (cf.\ \cite{hugonnier.kramkov.04, delbaen.schachermayer.98} for more details). 

\begin{defi}
A strategy $\varphi \in \Theta(v)$ is called \textit{maximal}\index{trading strategy!maximal} if the terminal value $v+\varphi\mal S_T$ of its wealth process is not dominated by that of any other strategy in $\Theta(v)$. An arbitrary strategy $\varphi$ is called \textit{acceptable} if its wealth process can be written as
$$ v+\varphi \mal S= v'+\varphi' \mal S -(v''+\varphi'' \mal S)$$
for some $v',v'' \in \rp$ and $\varphi',\varphi'' \in L(S)$ such that $v'+\varphi'\mal S \geq 0$, $v''+\varphi''\mal S \geq 0$ and, in addition, $\varphi''$ is maximal. For $v \in (0,\infty)$ and $q \in \rr$ we denote by 
$$ \Theta^q(v):=\{ \varphi \in L(S): \mbox{$\varphi$ is acceptable, } v+\varphi \mal S_T-qH \geq 0\},$$
the set of acceptable strategies whose terminal value dominates $qH$.
\end{defi}

\begin{bem}
Given Assumption \ref{a:1}, we have $\Theta(v)=\Theta^0(v)$ by \cite[Theorem 5.7]{delbaen.schachermayer.98} combined with \cite[Lemma 3.1 and Proposition 3.1]{kallsen.03} .
\end{bem}

Let an initial endowment of $v \in (0,\infty)$ be given. If the investor sells $q$ units of $H$ for a price of $x \in \rr$ each, her initial position consists of $v+qx$ in cash as well as $-q$ units of the contingent claim $H$. Hence $\Theta^q(v+qx)$ represents the natural set of admissible trading strategies for utility functions defined on $\rp$. The maximal expected utility the investor can achieve by dynamic trading in the market is then given by
\begin{equation*}
U^q(v+qx):=\sup_{\varphi \in \Theta^q(v+qx)}E(u(v+qx+\varphi \mal S_T-qH)).
\end{equation*}

\begin{defi}\index{utility indifference price}
Fix $q \in \rr$. A number $\pi^q \in \rr$ is called \textit{utility indifference price} of $H$ if 
\begin{equation}\label{e:indiff}
U^q(v+q\pi^q)=U(v).
\end{equation}
\end{defi}

Existence of indifference prices does not hold in general for power utility. However, a unique indifference price $\pi^q$ always exists if the number $q$ of contingent claims sold is sufficiently small or, conversely, if the initial endowment $v$ is sufficiently large.

\begin{lemma} \label{l:exindiffprice}
Suppose Assumptions \ref{a:1}, \ref{a:2} and \ref{a:3} hold. Then a unique indifference price exists for sufficiently small $q$. More specifically, \eqref{e:indiff} has a unique solution $\pi^q$ if $q < \frac{v}{2w}$, respectively if $q < \frac{v}{w}$ and $H \geq 0$, where $w$ denotes the initial endowment of the superhedging strategy for $H$ from Assumption \ref{a:3}.
\end{lemma}

\bpf First notice that $g^q_v:x \mapsto U^q(v+qx)$ is concave and strictly increasing on its effective domain. By \cite[Theorem 2.1]{kramkov.schachermayer.99}, $g^q_v(x) \leq U(v+qx+qw)<\infty$ for all $x \in \rr$.  For $H \geq 0$ and $q < \frac{v}{w}$ we have $g^q_v(x)>-\infty$ for $x > w-\frac{v}{q}$. In particular, $g^q_v$ is continuous and strictly increasing on $(w-\frac{v}{q},\infty)$ and in particular on $[0,w]$ by \cite[Theorem 10.1]{rockafellar.70}. By $H \geq 0$ we have $g^q_v(0) \leq U(v)$. Moreover, Assumption \ref{a:3} implies $g^q_v(w) \geq  U(v)$. Hence there exists a unique solution $\pi^q \in [0,w]$ to $g^q_v(x)=U(v)$. Similarly, for general $H$ and $q<\frac{v}{2w}$, the function $g^q_v$ is finite, continuous and strictly increasing on an open set containing $[-w,w]$. Moreover, $g^q_v(-w)\leq U(v)$ and $g^q_v(w) \geq U(v)$. Hence there exists a unique $\pi^q \in (-w,w)$ such that $g^q_v(\pi^q)=U(v)$. This proves the assertion. \ep\\

We now turn to optimal trading strategies in the presence of random endowment. Their existence has been established by \cite{cvitanic.al.99b} resp.\ \cite{hugonnier.kramkov.04} in the bounded resp.\ general case.

\begin{satz}\label{t:exindiffhedge}  
Fix $q \in \rr$ satisfying the conditions of Lemma \ref{l:exindiffprice} and suppose Assumptions \ref{a:1}, \ref{a:2} and \ref{a:3} are satisfied. Then there exists $\varphi^q \in \Theta^q(v+q\pi^q)$ such that
$$ E(u(v+q \pi^q+\varphi^q  \mal S_T-qH))=U^q(v+q\pi^q).$$
Moreover, the corresponding optimal value process $v+q\pi^q+\varphi^q \mal S$ is unique.
\end{satz}    

\bpf This follows from \cite[Theorem 2 and Corollary 1]{hugonnier.kramkov.04} because the proof of Lemma \ref{l:exindiffprice} shows that $(v+q\pi^q,q)$ belongs to the interior of $\{(x,r) \in \rr^2:\Theta^r(x) \neq \emptyset\}$.  \ep\\

Without contingent claims, the investor will trade according to the strategy $\widehat{\varphi}$, whereas she will invest into $\varphi^q$ if she sells $q$ units of $H$ for $\pi^q$ each. Hence, the difference between both strategies represents the action the investors needs to take in order to compensate for the risk of selling $q$ units of $H$. This motivates the following notion:

\begin{defi}
The trading strategy $\varphi^q-\widehat{\varphi}$ is called \emph{utility-based hedging strategy}.
\end{defi}

\section{The asymptotic results of Kramkov and S\^irbu}\label{s:kramkov.sirbu}\setcounter{equation}{0}
We now give a brief exposition of some of the deep results of \cite{kramkov.sirbu.06, kramkov.sirbu.06b} concerning the existence and characterization of first-order approximations of utility-based prices and hedging strategies in the following sense.

\begin{defi}\label{d:marginalprice}
Real numbers $\pi^0$ and $\pi'$ are called \textit{marginal utility-based price}\index{marginal utility-based price} resp.\ \emph{risk premium}\index{risk premium} per option sold if
$$ \pi^q= \pi^0+q\pi'+o(q)$$
for $q \to 0$, where $\pi^q$ is well-defined for sufficiently small $q$ by Lemma \ref{l:exindiffprice}. A trading strategy $\varphi' \in L(S)$ is called \textit{marginal utility-based hedging strategy}\index{hedging strategy!marginal utility-based} if there exists $v' \in \rr$ such that
\begin{equation}\label{eq:hedge}
 \lim_{q \to 0}\frac{(v+q\pi^q+\varphi^q\mal S_T)-(v+\widehat{\varphi}\mal S_T)-q(v'+\varphi'\mal S_T)}{q}=0
 \end{equation}
in $P$-probability and $(v'+\varphi'\mal S)\widehat{Y}$ is a martingale for the dual minimizer $\widehat{Y}$ of the pure investment problem.
\end{defi}

\begin{bem}
\cite[Theorems A.1, 8, and 4]{kramkov.sirbu.06} show that for power utility functions, a trading strategy $\varphi'$ is a marginal utility-based hedging strategy in the sense of Definition \ref{d:marginalprice} if and only if it is a marginal hedging strategy in the sense of \cite[Definition 2]{kramkov.sirbu.06b}. 
\end{bem}

The asymptotic results of \cite{kramkov.sirbu.06,kramkov.sirbu.06b} are derived subject to two technical assumptions.

\begin{assumption}\label{a:4}
The following process is \emph{$\sigma$-bounded}:
$$S^{\$}:= \left( \frac{1}{\E(-\til{a}\mal S)},\frac{S}{\E(-\til{a}\mal S)}\right).$$ 
\end{assumption}
The reader is referred to \cite{kramkov.sirbu.06} for more details on $\sigma$-bounded processes as well as for sufficient conditions that ensure the validity of this assumption. In our concrete examples, we have the following:

\begin{lemma}\label{lem:levybound}
\begin{enumerate}
\item Let $S=S_0\E(X)>0$ for a non-monotone L\'evy process $X$ with finite $(1-p)$-th moments. Then Assumption \ref{a:4} holds if the optimizer $\widehat{\eta}$ from Example \ref{bsp:levyinvest} lies in the \emph{interior} of the set $\scr{C}^0$ of fractions of wealth in stocks leading to nonnegative wealth processes.
\item Assumption \ref{a:4} is automatically satisfied if the stock price $S$ is continuous. In particular, it holds in the BNS model.
\end{enumerate}
\end{lemma}

\bpf First consider Assertion 1. In view of \cite[Lemma 8]{kramkov.sirbu.06}, it suffices to show that $S^{\$}$ is bounded by a predictable process. If $\widehat{\eta} \geq 0$, there exists $\overline{\eta} \in \scr{C}^0$ with $\widehat{\eta}<\overline{\eta}$; hence $\Delta X>-1/\overline{\eta}$ by definition of $\scr{C}^0$. Consequently, $\widehat{\eta}\Delta X\geq-\widehat{\eta}/\overline{\eta}>-1$ and thus
$$ \left|\frac{1}{\E(-\til{a}\mal S)}\right|=\frac{1}{|\E(\widehat{\eta}X)|}=\frac{1}{|1+\widehat{\eta}\Delta X|}\frac{1}{|\E(\widehat{\eta}X)_{-}|} \leq \frac{\overline{\eta}}{(\overline{\eta}-\widehat{\eta})|\E(\widehat{\eta}X)_{-}|} \quad \mbox{a.s.},$$
which shows that the first component of $S^{\$}$ is bounded by a predictable process and hence $\sigma$-bounded. Likewise, if $\widehat{\eta}<0$, there exists $\underline{\eta} \in \scr{C}^0$ with $\underline{\eta}<\widehat{\eta}$. Then $\Delta X<-1/\underline{\eta}$ and in turn $\widehat{\eta}\Delta X>-\widehat{\eta}/\underline{\eta}>-1$. Hence it follows as above that $|1/\E(-\til{a}\mal S)|$ is bounded by a predictable process. The assertion for the second component of $S^{\$}$ follows similarly.

If the stock price process is continuous, both $S$ and $\E(-\til{a} \mal S)$ are predictable. Hence Assertion 2 follows immediately from \cite[Lemma 8]{kramkov.sirbu.06}. \ep \\

Since $\E(-\til{a}\mal S)\widehat{Y}$ is a martingale with terminal value $\E(-\til{a}\mal S)_T^{1-p}$, we can define an equivalent probability measure $Q^\$ \sim P$ via
\begin{equation*}\label{e:qdollar}
\frac{dQ^\$}{dP}:=\frac{\E(-\til{a}\mal S)_T^{1-p}}{C_0}, \quad C_0:=E(\E(-\til{a}\mal S)_T^{1-p}).
\end{equation*}
Let $\H^2_0(Q^\$)$ be the space of square-integrable $Q^{\$}$-martingales starting at $0$ and set
\begin{equation}\label{eq:md2}
\scr{M}_\$^2:= \left\{ M \in \H^2_0(Q^\$): M= \varphi \mal S^\$ \mbox{ for some } \varphi \in L(S^{\$})\right\}.
\end{equation}

\begin{assumption}\label{a:5}
There exists a constant $w^\$\geq 0$ and a process $M^\$ \in \scr{M}_\$^2$, such that
$$ |H^\$| \leq w^\$+M^\$_T$$
for 
$$H^\$:=\frac{H}{\E(-\til{a}\mal S)_T}.$$
\end{assumption}

Assumption \ref{a:5} means that the claim under consideration can be superhedged with portfolios as in \eqref{eq:md2}. Note that this is again evidently satisfied for European puts and calls.

\begin{bem}
By \cite[Remark 1]{kramkov.sirbu.06}, Assumption \ref{a:5} implies that Assumption \ref{a:3} holds. In particular,  it ensures that indifference prices and utility-based hedging strategies exist for sufficiently small $q$ if the pure investment problem is well-posed, i.e., if Assumptions \ref{a:1} and \ref{a:2} are also satisfied.
\end{bem}

In the proof of \cite[Lemma 1]{kramkov.sirbu.06b} it is shown that the process
$$ V^\$_t:= E_{Q^\$}\left(H^\$| \scr{F}_t \right), \quad  t\in [0,T]$$
is a square-integrable $Q^\$$-martingale. Hence it admits a decomposition
\begin{equation}\label{e:KW}
V^\$ = E_{Q^\$}\left(H^\$\right)+\xi \mal S^\$+N^\$=\frac{1}{C_0}E\left(\E(-\til{a}\mal S)_T^{-p} H\right)+\xi \mal S^\$+N^\$, 
\end{equation}
where $\xi \mal S^\$ \in \scr{M}_\$^2$ and $N^\$$ is an element of the orthogonal complement of $\scr{M}_\$^2$ in $\H^2_0(Q^\$)$. Note that this decomposition coincides with the classical \emph{Galtchouk-Kunita-Watanabe} decomposition if $S^\$$ itself is a square-integrable martingale. The following theorem is a reformulation of the results of \cite{kramkov.sirbu.06,kramkov.sirbu.06b} applied to power utility, and also contains a feedback representation of the utility-based hedging strategy in terms of the original numeraire.

\begin{satz}\label{s:KS}
Suppose Assumptions \ref{a:1}, \ref{a:2}, \ref{a:4}, and \ref{a:5} hold. Then the marginal utility-based price $\pi^0$ and the risk premium $\pi'$ exist and are given by
\begin{equation*}
\pi^0		= \frac{1}{C_0}E(\E(-\til{a}\mal S)_T^{-p} H), \quad \pi'= \frac{p}{2v}E_{Q^\$}((N^\$_T)^2).
\end{equation*}
A marginal-utility-based hedging strategy $\phi'$ is given in feedback form as the solution of the stochastic differential equation
\begin{equation*}
\phi'	= (\widetilde{a},E_d+\widetilde{a}S_{-}^{\top})\xi-\left(\pi^0+\phi' \mal S_{-}\right)\widetilde{a},
\end{equation*} 
with $\xi$ from \eqref{e:KW}, and where $E_d$ denotes the identity matrix on $\mathbb{R}^d$.
\end{satz}

\bpf The first two assertions follow immediately from \cite[Theorems A.1, 8, and 4]{kramkov.sirbu.06} adapted to the present notation. For the third, \cite[Theorem 2]{kramkov.sirbu.06b} and \cite[Theorems A.1, 8, and 4]{kramkov.sirbu.06} yield 
\begin{equation}\label{e:Convergence}
\lim_{q \to 0}\frac{(v+q\pi^q+\varphi^q \mal S_T)-(v+\widehat{\varphi} \mal S_T)-q\E(-\til{a}\mal S)_T(\pi^0+\xi \mal S^\$_T)}{q}=0.
\end{equation}
because the process $X_T'(x)$ from \cite[Equation (23)]{kramkov.sirbu.06b} coincides with $\E(-\til{a} \mal S)$ for power utility. Set $$\xi^0:=\pi^0+\xi \mal S^\$-\xi^{\top}S^\$=\pi^0+\xi \mal S^\$_{-}-\xi^{\top}S^\$_{-}.$$ 
Then we have $(\xi^0,\xi^2,\ldots,\xi^{d+1}) \in L((\E(-\til{a} \mal S),S))$ and 
\begin{equation}\label{e:GK}
\pi^0+(\xi^0,\xi^2,\ldots,\xi^{d+1}) \mal (\E(-\til{a}\mal S),S)=\E(-\til{a}\mal S)(\pi^0+\xi \mal S^{\$}) 
\end{equation}
by \cite[Proposition 2.1]{kallsen.goll.99a}. The predictable sets $D_n := \{|\til{a}|\leq n, |S_{-}|\leq n, |(\xi^0,\xi)|\leq n\}$ increase to $\Omega \times \rp$, the predictable process $(\til{a},E_d+\til{a}S_{-}^{\top})\xi 1_{D_n}$ is bounded, and we have
\begin{align*}
&((\til{a},E_d+\til{a}S_{-}^{\top})\xi 1_{D_n}) \mal S\\
\quad &= ((\E(-\til{a}\mal S)_{-}\xi^{\top}S^{\$}_{-}\til{a}+(\xi^2,\ldots,\xi^{d+1}))1_{D_n}) \mal S\\
\quad &= ((\xi^0,\xi^2,\ldots,\xi^{d+1})1_{D_n})\mal (\E(-\til{a}\mal S),S)+(\E(-\til{a}\mal S)_{-}(\pi^0+\xi \mal S^{\$}_{-})1_{D_n}) \mal (\til{a} \mal S)\\
\quad &= 1_{D_n}\mal ((\xi^0,\xi^2,\ldots,\xi^{d+1})\mal (\E(-\til{a}\mal S),S)+(\E(-\til{a}\mal S)_{-}(\pi^0+\xi \mal S^{\$}_{-})) \mal (\til{a} \mal S)).
\end{align*}
By \cite[Lemma 2.2]{kallsen.03} and \eqref{e:GK}, this implies $(\til{a},E_d+\til{a}S_{-}^{\top})\xi \in L(S)$ as well as
$$\pi^0+((\til{a},E_d+\til{a}S_{-}^{\top})\xi) \mal S =  \E(-\til{a}\mal S)(\pi^0+\xi \mal S^{\$})+(\E(-\til{a}\mal S)_{-}(\pi^0+\xi \mal S^{\$}_{-})) \mal (\til{a} \mal S).$$
Hence $\E(-\til{a} \mal S)(\pi^0+\xi \mal S^\$)$ solves the stochastic differential equation
\begin{equation}\label{e:SDE}
G=\pi^0+((\til{a},E_d+\til{a}S^\top_{-})\xi)\mal S-G_{-} \mal (\til{a} \mal S).
\end{equation}
By \cite[(6.8)]{jacod.79} this solution is unique. Since we have shown  $(\til{a},E_d+\til{a}S_{-}^{\top})\xi \in L(S)$ above, it follows as in the proof of \cite[Lemma 4.9]{cerny.kallsen.05} that $\phi'$ is well-defined. $\pi^0+\phi' \mal S$ also solves \eqref{e:SDE}, hence we obtain
$$\E(-\til{a}\mal S)(\pi^0+\xi \mal S^{\$})=\pi^0+\phi' \mal S.$$
In view of \eqref{e:Convergence}, the process $\pi^0+\phi' \mal S$ therefore satisfies \eqref{eq:hedge}, so that $\phi'$ is indeed a marginal utility-based hedge in the sense of Definition \ref{d:marginalprice}.\ep

\begin{bem}\label{rem:qopt}
If the dual minimizer $\widehat{Y}$ is a martingale and hence -- up to the constant $C_0$ -- the density process of the \emph{$q$-optimal martingale measure} $Q_0$ with respect to $P$, the generalized Bayes formula yields $V^{\$}_t=E_{Q_0}(H|\F_t)/ \E(-\til{a} \mal S)_t$. In particular, the marginal utility-based price of the claim $H$ is given by its expectation $\pi^0=E_{Q_0}(H)$ under $Q_0$ in this case. 
\end{bem}

The computation of the optimal strategy $\widehat{\varphi}$ and the corresponding dual minimizer $\widehat{Y}$ in the pure investment problem \ref{e:pureinvestment} has been studied extensively in the literature. In particular, these objects have been determined explicitly in a variety of Markovian models using stochastic control theory resp.\ martingale methods. Given $\scr{E}(-\til{a}\mal S)$, the computation of $\pi^0$ can then be dealt with using integral transform methods or variants of the Feynman-Kac formula. Consequently, we suppose from now on that $\widehat{\varphi}$ and $\pi^0$ are known and focus on how to obtain $\pi'$ and  $\varphi'$.

As reviewed above, \cite{kramkov.sirbu.06,kramkov.sirbu.06b} show that $\varphi'$ and $\pi'$ can be obtained by calculating the generalized Galtchouk-Kunita-Watanabe decomposition \eqref{e:KW}. Since $S^\$$ is generally only a $Q^\$$-supermartingale, this is typically very difficult. If however, $S^\$$ happens to be a square-integrable $Q^\$$-martingale, \eqref{e:KW} coincides with the classical Galtchouk-Kunita-Watanabe decomposition. By \cite{foellmer.sondermann.86}, this shows that $\xi$ represents the mean-variance optimal hedging strategy for the claim $H$ hedged with $S^\$$ under the measure $Q^\$$ and $E_{Q^{\$}}((N^\$_T)^2)$ is given by the corresponding minimal expected squared hedging error in this case. Moreover, $\xi$ and $E_{Q^\$}((N_T^{\$})^2)$ can then be characterized in terms of semimartingale characteristics.

\begin{assumption}\label{a:6}
$S^\$$ is a square-integrable $Q^\$$-martingale.
\end{assumption}

For exponential L\'evy models, this assumption satisfied if the budget constraint $\scr{C}^0$ is ``not binding'' for the optimal fraction $\widehat{\eta}$ of stocks and if, in addition, the driving L\'evy process is square-integrable. For the BNS model it is only a matter of integrability.

\begin{lemma}\label{lem:levymart}
\begin{enumerate}
\item Let $S=S_0\E(X)>0$ for a non-monotone L\'evy process $X$ with finite second moments. Then Assumption \ref{a:6} is satisfied if the optimizer $\widehat{\eta}$ of the pure investment problem lies in the interior of $\scr{C}^0$.
\item Let $S=S_0 \E(X)$, where $(y,X)$ is a BNS model. If $p>1$ or \eqref{eq:bnsint} holds, then $S^{\$}$ is a $Q^{\$}$-martingale.
\end{enumerate}
\end{lemma}

\bpf If $\widehat{\eta}$ lies in the interior of $\scr{C}^0$, it follows from \cite[Proposition 5.12]{nutz.09c} that the dual optimizer $Y=L\E(-\til{a}\mal S)^{-p}$ is a local martingale. Since it is also the exponential of a L\'evy process (cf.\ \cite[Section 6]{nutz.10c}), it is in a fact a true martingale. Thus it is -- up to normalization with $1/L_0$ -- the density process of the $q$-optimal martingale measure by \cite[Remark 5.18]{nutz.09c}. Combined with \cite[Proposition III.3.8]{js.03}, this yields that $S^{\$}$ is a $Q^{\$}$-martingale and it remains to show that $S^{\$}$ is square-integrable. By Propositions \ref{p:Ito}, \ref{p:Int}, and \ref{p:Measure}, the process $S^{\$}=(1/\E(\widehat{\pi}X),S_0\E(X)/\E(\widehat{\pi}X))$ is the stochastic exponential $(\E(R^1),\E(R^2))$ of a semimartingale $R$ with local $Q^{\$}$-characteristics 
$$\left(0, \begin{pmatrix} \widehat{\eta}^2 & -\widehat{\eta}(1-\widehat{\eta}) \\ -\widehat{\eta}(1-\widehat{\eta}) & (1-\widehat{\eta})^2 \end{pmatrix}, G \mapsto \int 1_{G}\left(\frac{-\widehat{\eta}x}{1+\widehat{\eta}x},\frac{(1-\widehat{\eta})x}{1+\widehat{\eta}x}\right)(1+\widehat{\eta}x)^{1-p}F^X(dx)\right),$$
relative to the truncation function $h(x)=x$ on $\mathbb{R}^2$. This truncation function can be used because $R$ is $Q^{\$}$-locally a square-integrable martingale. By \cite[Propositions II.2.29 and III.6.35]{js.03}, this holds because $X$ is square-integrable, $|1/(1+\widehat{\eta}\Delta X)|$ is bounded (cf.\ the proof of Lemma \ref{lem:levybound}) and hence $\int x_1^2 F^{R,\$}(dx)<C\int x^2 F^X(dx)<\infty$ and $\int x_2^2 F^{R,\$}(dx)<C \int x^2 F^X(dx)<\infty$ for some constant $C \in \rp$ (cf.\ \cite[Theorem II.1.8]{js.03}). As the $Q^{\$}$-characteristics of $R$ are deterministic, $R$ is a $Q^{\$}$-L\'evy process by \cite[Corollary II.4.19]{js.03} and a square-integrable martingale by \cite[Proposition I.4.50]{js.03}. Therefore $S^{\$}=(\E(R^1),\E(R^2))$ is a square-integrable martingale as well by \cite[Lemma A.1.(x)]{muhlekarbe.nutz.10}. This proves Assertion 1.

Assertion 2 is shown in the proof of \cite[Theorem 3.3]{kallsen.muhlekarbe.10}.
\ep \\

The square-integrability of $S^{\$}$ in the BNS model is discussed in Remarks \ref{rem:bnssquare1} and \ref{rem:bnssquare} below. Given Assumption \ref{a:6}, we have the following representation.

\begin{lemma}\label{l:Char}
Suppose Assumptions \ref{a:1}, \ref{a:2}, \ref{a:4}, \ref{a:5} and \ref{a:6} hold. Denote by $\tilde{c}^{(S^{\$},V^{\$})\$}$ the modified second $Q^{\$}$-characteristic of $(S^{\$},V^{\$})$ with respect to some $A \in \apl$ (cf.\  Appendix \ref{appendix}). Then 
\begin{gather}
\xi=(\tilde{c}^{S^\$ \$})^{-1}\tilde{c}^{S^\$,V^\$ \$},\label{e:xi}\\
E_{Q^\$}((N^\$_T)^2)=E_{Q^\$}\left((\tilde{c}^{V^\$ \$}-(\tilde{c}^{S^\$,V^\$ \$})^{\top}(\tilde{c}^{S^\$ \$})^{-1}\tilde{c}^{S^\$,V^\$ \$})\mal A_T\right).\notag
\end{gather}
\end{lemma}

\bpf Since $S^{\$}$ is a square integrable $Q^{\$}$-martingale by Assumption \ref{a:6}, the claim follows from \cite[Theorems 4.10 and 4.12]{cerny.kallsen.05} applied to the martingale case. \ep

\section{An alternative representation}\label{s:alternative}\setcounter{equation}{0}

We now develop our alternative representation of power utility-based prices and hedging strategies. As explained in the introduction, they can -- morally speaking -- be represented as the solution to a mean-variance hedging problem relative to the original numeraire, but subject to yet another probability measure $P^{\eur} \sim P$. Given Assumption \ref{a:6}, the latter can be defined as follows: 

\begin{equation*}
\frac{dP^{\eur}}{dP}:=\frac{\E(-\til{a}\mal S)^{-1-p}_T}{C_1}, \quad C_1:=E(\E(-\til{a} \mal S)_T^{-1-p}).
\end{equation*}

\begin{bem}\label{bem:Leuro}
If we write the density process of $P^{\eur}$ with respect to $P$ as $L^{\eur}\E(-\til{a} \mal S)^{-1-p}/C_1$ for a semimartingale $L^{\eur}>0$ with $L^{\eur}_T=1$, the local joint $P$-characteristics of $S$ and $K^{\eur}:=\L(L^{\eur})$ relative to some truncation function $(h_1,h_2)$ on $\rr^{d} \times \rr$ satisfy 
\begin{equation}\label{e:integrable}
\int_{\{|x|>1\}} (1+x_2)(1-\til{a}^{\top}x_1)^{-1-p}F^{(S,K^{\eur})}(dx)<\infty,
\end{equation}
and solve
\begin{align}
0 &= b^{K^{\eur}}+(1+p)\til{a}^{\top}b^S+(1+p)\til{a}^{\top}c^{S,K^{\eur}}+\frac{(p+1)(p+2)}{2}\til{a}^{\top}c^{S}\til{a}\label{e:drift}\\
 &  \quad +\int\left((1+x_2)(1-\til{a}^{\top}x_1)^{-1-p}-1-h_2(x_2)-(1+p)\til{a}^{\top}h_1(x_1)\right)F^{(S,K^{\eur})}(dx),\notag
\end{align}  
by \cite[Lemma 3.1]{kallsen.03} and Propositions \ref{p:Int}, \ref{p:Ito} . Conversely, if a strictly positive semimartingale $L^{\eur}=L^{\eur}_0\E(K^{\eur})$ satisfies $L^{\eur}_T=1$ and \eqref{e:integrable}, \eqref{e:drift}, then $L^{\eur}\E(-\til{a} \mal S)^{-1-p}/C_1$ is a $\sigma$-martingale and the density process of $P^{\eur}$ if it is a true martingale. 
\end{bem}

In concrete models, the drift condition \eqref{e:drift} often allows to determine $L^{\eur}$ by making an appropriate parametric ansatz. For exponential L\'evy models and the BNS model, this leads to the following results.

\begin{bsp}\label{bsp:levyleur}
\begin{enumerate}
\item For exponential L\'evy models as in Example \ref{bsp:levyleur}, plugging the ansatz $a^{\eur}t$ with $a^{\eur} \in \mathbb{R}$ for $K^{\eur}$ into \eqref{e:drift} yields
\begin{align*}
a^{\eur}=&(1+p)\widehat{\eta}b^X-\frac{(p+1)(p+2)}{2} \widehat{\eta}^2 c^X\\
&\qquad -\int\left((1+\widehat{\eta}x)^{-1-p}-1+(1+p)\widehat{\eta}h(x)\right)F^X(dx).
\end{align*}
This expression is well-defined because the integrand is of order $O(x^2)$ for small $x$ and bounded on the support of $F^X$ by the proof of Lemma \ref{lem:levybound} and \cite[Theorem II.1.8]{js.03}.
One then easily verifies that $L^{\eur}_t=\exp(a^{\eur}(T-t))$. Indeed,  the strictly positive $\sigma$-martingale $L^{\eur}\E(\widehat{\eta}X)$ is a true martingale because it is also the stochastic exponential of a L\'evy process.
\item \label{bnsleur} For the BNS model, one has to make a more general ansatz for $L^{\eur}_t$. Choosing $\exp(\alpha_0^{\eur}(t)+\alpha_1^{\eur}(t)y_t)$ with smooth functions $\alpha^{\eur}_0, \alpha^{\eur}_1$ satisfying $\alpha^{\eur}_0(T)=\alpha^{\eur}_1(T)=0$ as in \cite{kallsen.muhlekarbe.10}, insertion into \ref{e:drift} leads to
\begin{align*}
\alpha_1^{\eur}(t)=\frac{(1+p)(2-p)}{2p^2} \mu^2 \frac{1-e^{-\lambda (T-t)}}{\lambda}, \quad \alpha_0^{\eur}(t)=\int_t^T \psi^{Z}(\alpha_1^{\eur}(s))ds.
\end{align*}
Then, \eqref{e:integrable} is satisfied and $\exp(\alpha^{\eur}_0(t)+\alpha^{\eur}_1(t)y)\E(\widehat{\eta}X)^{-1-p}$ is a $\sigma$-martingale if 
\begin{equation}\label{eq:intint}
\int_{\{|x|>1\}}e^{\alpha_1^{\eur}(t)z}F^Z(dz)<\infty \quad \mbox{ for all } t \in [0,T].
\end{equation}
\eqref{eq:intint} automatically holds for $p \geq 2$, because $\alpha_1^{\eur} \leq 0$ in this case. For $p \in (0,2)$, \eqref{eq:intint} is satisfied if
\begin{equation}\label{eq:leurint}
\int_1^\infty \exp\left(\frac{(1+p)(2-p)}{2p^2}\mu^2 \frac{1-e^{-\lambda T}}{\lambda} z\right)F^Z(dz)<\infty.
\end{equation}
In either case, the true martingale property of the exponentially affine $\sigma$-martingale $\exp(\alpha^{\eur}_0(t)+\alpha^{\eur}_1(t)y)\E(\widehat{\eta}X)^{-1-p}$ follows from \cite[Corollary 3.9]{kallsen.muhlekarbe.10b}. This shows that $L^{\eur}$ is indeed given by $\exp(\alpha^{\eur}_0(t)+\alpha^{\eur}_1(t)y)$.
\end{enumerate}
\end{bsp}

\begin{bem}\label{rem:bnssquare1}
Part 2 of Example \ref{bsp:levyleur} shows that in the BNS model the first component of $S^{\eur}$ is square-integrable if $p \geq 2$ or \eqref{eq:leurint} holds. Hence the measure $P^{\eur}$ is well defined with density process $L^{\eur}\E(\widehat{\eta}X)^{-1-p}$ in either case.
\end{bem}

As motivated in the introduction, the measures $P^{\eur}$ and $Q^{\$}$ are linked as follows.

\begin{lemma}\label{l:densityopp}
Suppose Assumptions \ref{a:1}, \ref{a:2} and \ref{a:6} hold. Then the process
$$ L^{\$}_t := E_{P^{\eur}}\left(\frac{\E(-\til{a}\mal S)_T^2}{\E(-\til{a}\mal S)_t^2} \bigg| \scr{F}_t\right), \quad 0 \leq t \leq T,$$
satisfies $L^{\$}_T=1$ and the density process of $Q^{\$}$ with respect to $P^{\eur}$ is given by
$$ E_{P^{\eur}}\left(\frac{dQ^\$}{dP^{\eur}}\bigg| \scr{F}_t\right)= \frac{C_1}{C_0} L^{\$}_t \E(-\til{a} \mal S)_t^2= \frac{L^{\$}_t \E(-\til{a} \mal S)_t^2}{L^{\$}_0}.$$
In particular, $L^{\$}, L^{\$}_{-}>0$ and the stochastic logarithm $K^{\$}:=\scr{L}(L^{\$})$ is  well-defined.
\end{lemma}

\bpf The first part of the assertion is trivial, whereas the second follows from $dQ^{\$}/dP^{\eur}=\frac{C_1}{C_0}\E(-\til{a}\mal S)^2_T$. Since $\E(-\til{a}\mal S),\E(-\til{a}\mal S)_{-}>0$, \cite[I.2.27]{js.03} yields $L^{\$}, L^{\$}_{-}>0$ and hence the third part of the assertion by \cite[II.8.3]{js.03}. \ep 

\begin{bem}\label{bem:link}
$L^{\$}$ is linked to the opportunity process $L$ of the pure investment problem and the process $L^{\eur}$ from Remark \ref{bem:Leuro} via
$$L_0^{\$}\E(K^{\$})=L^{\$}=\frac{L}{L^{\eur}}=\frac{L_0\E(K)}{L^{\eur}_0\E(K^{\eur})},$$
by the generalized Bayes' formula, $L_T=L^{\eur}_T=1$, and because $L\E(-\til{a}\mal S)^{1-p}$ as well as $L^{\eur}\E(-\til{a}\mal S)^{-1-p}$ are martingales. 
\end{bem}

In our examples, this leads to the following.

\begin{bsp}\label{bsp:levyld}
\begin{enumerate}
\item Suppose $S=S_0 \E(X)>0$ for a non-monotone L\'evy process with finite second moments. Then $L^{\$}=\exp((a-a^{\eur})(T-t))$ and $K_t^{\$}=(a^{\eur}-a)t$ for $a$ and $a^{\eur}$ as in Examples \ref{bsp:levyinvest} resp.\ \ref{bsp:levyleur}.
\item Let $S=S_0 \E(X)>0$ for a BNS model satisfying \eqref{eq:leurint} if $p<2$ and, additionally, \eqref{eq:bnsint} if $p<1$. Then $L^{\$}_t=\exp(\alpha_0^{\$}(t))+\alpha_1^{\$}(t)y_t)$ and, by It\^o's formula, 
$$K^{\$}_t= \alpha_0^{\$}(t)-\alpha_0^{\$}(0)+\alpha_1^{\$}(t)y_t-\alpha_1^{\$}(0)y_0+\sum_{s \leq t}\left(e^{\alpha_1^{\$}(s)\Delta Z_s}-1-\alpha_1^{\$}(s)\Delta Z_s\right),$$
for $\alpha_i^{\$}=\alpha_i-\alpha_i^{\eur}$, $i=0,1$, with $\alpha_i$ and $\alpha_i^{\eur}$ as in Examples \ref{bsp:levyinvest} resp.\ \ref{bsp:levyleur}.
\end{enumerate}
\end{bsp}

Now define 
\begin{equation*}
V_t:=\E(-\til{a}\mal S)_t V^\$_t=\frac{E(\E(-\til{a}\mal S)_T^{-p} H|\scr{F}_t)}{L_t \E(-\til{a}\mal S)_t^{-p}}, \quad 0 \leq t \leq T,
\end{equation*}
which coincides with the conditional expectation under the $q$-optimal martingale measure $Q_0$, if the latter exists. Denote by
\begin{equation*}
\left(\begin{pmatrix} b^{S\eur} \\ b^{V\eur} \\ b^{K^{\$}\eur} \end{pmatrix}, \begin{pmatrix} c^{S\eur} & c^{S,V\eur} & c^{S,K^{\$}\eur} \\ c^{V,S\eur} & c^{V\eur} & c^{V,K^{\$}\eur}\\ c^{K^{\$},S\eur} & c^{K^{\$},V\eur} & c^{K^{\$}\eur} \end{pmatrix}, F^{(S,V,K^{\$})\eur},A\right)
\end{equation*}
$P^{\eur}$-differential characteristics of the semimartingale $(S,V,K^{\$})$ and define
\begin{eqnarray*}
\tilde{c}^{S\star}&:=& \frac{1}{1+\Delta A^{K^{\$}}} \left(c^{S\eur}+\int(1+x_3)x_1 x_1^{\top} F^{(S,V,K^{\$})\eur}(dx)\right),\\
\tilde{c}^{S,V\star}&:=&\frac{1}{1+\Delta A^{K^{\$}}} \left(c^{S,V\eur}+\int(1+x_3)x_1 x_2 F^{(S,V,K^{\$})\eur}(dx)\right), \\
\tilde{c}^{V\star}&:=& \frac{1}{1+\Delta A^{K^{\$}}} \left(c^{V\eur}+\int(1+x_3)x_2^2 F^{(S,V,K^{\$})\eur}(dx)\right),
\end{eqnarray*}
where $K^{\$}=K^{\$}_0+A^{K^{\$}}+M^{K^{\$}}$ denotes an arbitrary $P^{\eur}$-semimartingale decomposition of $K^{\$}$. We then have the following representation of the marginal utility-based hedging strategy  $\varphi'$ and the risk premium $\pi'$ in terms of semimartingale characteristics, which is the main result of this paper.

\begin{satz}\label{s:mainresult}
Suppose Assumptions \ref{a:1}, \ref{a:2}, \ref{a:4}, \ref{a:5} and \ref{a:6} hold. Then $\tilde{c}^{S\star}, \tilde{c}^{S,V\star}, \tilde{c}^{V\star}$ are well-defined, the strategy $\varphi'$ given in feedback form as the solution of the stochastic differential equation 
\begin{equation*}\label{e:marginalhedge}
\varphi'=(\tilde{c}^{S\star})^{-1}\tilde{c}^{S,V\star}-\left(\pi^0+\varphi' \mal S_{-}-V_{-}\right)\widetilde{a}
\end{equation*}
is a marginal utility-based hedge, and the corresponding risk premium is 
\begin{equation*}
\pi'=\frac{pC_1}{2vC_0}E_{P^{\eur}}\bigg(\big(\left(\tilde{c}^{V\star}-(\tilde{c}^{S,V\star})^{\top}(\tilde{c}^{S\star})^{-1}\tilde{c}^{S,V\star}\right)L^{\$}\big) \mal A_T\bigg).
\end{equation*}
\end{satz} 

\begin{bem}
As is customary for mean-variance optimal hedges \cite{schweizer.94,cerny.kallsen.05}, the strategy $\varphi'$ is described in ``feedback form'', i.e., it is computed as the solution of a stochastic differential equation involving its past trading gains $\varphi' \mal S_{-}$, which reduces to a simple recursive formula in discrete time (cf., e.g., \cite[Theorem 2.4]{schweizer.95}). Alternatively, the corresponding linear stochastic differential equation for $\varphi' \mal S$ can be solved \cite[Corollary 4.11]{cerny.kallsen.05},
$$\varphi' \mal S=\E(\-\til{a}\mal S) \left(\frac{(\tilde{c}^{S\star})^{-1}\tilde{c}^{S,V\star}+(V_{-}-\pi^0)\til{a}}{\E(-\til{a}\mal S)_{-}}\mal\left(S+\frac{\til{a}}{1-\til{a}^{\top} \Delta S} \mal [S,S]\right)\right),$$
leading to a cumbersome but explicit expression for the hedge $\varphi'$.
\end{bem}

In view of \cite[Theorems 4.10 and 4.12]{cerny.kallsen.05}, Theorem \ref{s:mainresult} states that the first-order approximations for $\varphi^q$ and $\pi^q$ can \textit{essentially} be computed by solving the mean-variance hedging problem for the claim $H$ under the (non-martingale) measure $P^{\eur}$ relative to the original numeraire. However, this assertion only holds true \textit{literally} if the dual minimizer $\widehat{Y}$ is a martingale and if the optimal strategy $\widehat{\varphi}$ in the pure investment problem is \textit{admissible} in the sense of \cite[Corollary 2.5]{cerny.kallsen.05}, i.e., if $\widehat{\varphi} \mal S_T \in L^2(P^{\eur})$ and $(\widehat{\varphi} \mal S)Z^Q$ is a $P^{\eur}$-martingale for any absolutely continuous signed $\sigma$-martingale measure $Q$ with density process $Z^Q$ and $\frac{dQ}{dP^{\eur}} \in L^2(P^{\eur})$. More precisely, in this case the strategy $-\widetilde{a}1_{\zu \tau,T\zu}\E(-\til{a}1_{\zu \tau,T\zu}\mal S)_{-}$ is \textit{efficient} on the stochastic interval $\zu \tau, T\zu$ in the sense of \cite[Section 3.1]{cerny.kallsen.05} and $\til{a}$ is the corresponding \emph{adjustment process} in the sense of \cite[Definition 3.8]{cerny.kallsen.05}. By \cite[Corollary 3.4]{cerny.kallsen.05} this in turn implies that $L^{\$}$ is the \textit{opportunity process} in the sense of \cite[Definition 3.3]{cerny.kallsen.05}. Hence it follows along the lines of \cite[Lemma 3.15]{cerny.kallsen.05} that the \textit{opportunity neutral measure} $P^\star$ with density process 
$$ Z^{P^\star}:=\frac{L^{\$}}{L^{\$}_0\scr{E}(A^{K^{\$}})}$$ 
exists. By \cite[Lemma 3.17 and Theorem 4.10]{cerny.kallsen.05}, $\tilde{c}^{S\star}, \tilde{c}^{V\star}, \tilde{c}^{S,V\star}$ indeed coincide with the corresponding modified second characteristics of $(S,V,K)$ under $P^\star$. Hence \cite[Theorems 4.10 and 4.12]{cerny.kallsen.05} yield that relative to the probability measure $P^{\eur}$, the process $\varphi'$ represents a variance-optimal hedging strategy for $H$ while the minimal expected squared hedging error of $H$ is given by the $2 C_0 v/(p C_1)$-fold of $\pi'$. Moreover, $V_t=E_{Q_0}(H|\scr{F}_t)$ and in particular the marginal utility-based price $\pi^0=E_{Q_0}(H)$ are given as conditional expectations under the \emph{variance-optimal martingale measure} $Q_0$ with respect to $P^{\eur}$, which coincides with the $q$-optimal martingale measure with respect to $P$. \\

{\sc Proof of Theorem \ref{s:mainresult}.}\ An application of Propositions \ref{p:Int} and \ref{p:Ito} yields the $P^{\eur}$-differential characteristics of the process $(S,V,\E(-\til{a}\mal S),\scr{L}(\frac{C_1}{C_0}L^{\$}\E(-\til{a}\mal S)^2))$. Now, since $\frac{C_1}{C_0}L^{\$} \E(-\til{a}\mal S)^2$ is the density process of $Q^\$$ with respect to $P^{\eur}$, the $Q^\$$-characteristics of $(S,V,\E(-\til{a}\mal S))$ can be obtained with Proposition \ref{p:Measure}. Another application of Proposition \ref{p:Ito} then allows to compute the $Q^\$$-characteristics of $(S^\$,V^\$)$. 

Since $S^\$ \in \H^2(Q^{\$})$ by Assumption \ref{a:6} and $V^\$ \in \H^2(Q^{\$})$ by the proof of \cite[Lemma 1]{kramkov.sirbu.06b}, the modified second characteristics $\tilde{c}^{V^\$ \$}$, $\tilde{c}^{S^\$,V^\$ \$}$ and $\tilde{c}^{S^\$ \$}$ exist and are given by
\begin{eqnarray}
\tilde{c}^{S^\$ \$}&=&\frac{1+\Delta A^{K^\$}}{\E(-\til{a}\mal S)_{-}^2}\begin{pmatrix} \widetilde{a}^{\top} \tilde{c}^{S\star} \til{a} &  \til{a}^{\top}\tilde{c}^{S\star}R^{\top} \\ R\tilde{c}^{S\star} \til{a} & R\tilde{c}^{S\star}R^{\top}\end{pmatrix}\label{e:tilcs},\\
\tilde{c}^{S^\$,V^\$ \$}&=&\frac{1+\Delta A^{K^\$}}{\E(-\til{a}\mal S)_{-}^2}\begin{pmatrix} \widetilde{a}^{\top} \\ R \end{pmatrix}\left(\tilde{c}^{S,V\star}+\tilde{c}^{S\star}\widetilde{a} V_{-}\right), \label{e:tilcsv}\\
\tilde{c}^{V^\$ \$}&=& \frac{1+\Delta A^{K^\$}}{\E(-\til{a}\mal S)_{-}^2}\left(\tilde{c}^{V\star}+2 V_{-}\widetilde{a}^{\top}\tilde{c}^{S,V\star}+V_{-}^2\til{a}^{\top} \til{c}^{S\star}\til{a}\right)\label{e:tilcv}
\end{eqnarray}
for $R:=E_d+S_{-}\til{a}^{\top}$. In particular it follows that $\tilde{c}^{V\star}$, $\tilde{c}^{S,V\star}$ and $\tilde{c}^{S\star}$ are well defined. By the definition of $\xi$ in Equation \eqref{e:xi} and \cite[Theorem 9.1.6]{albert.72} we have
$$\tilde{c}^{S^{\$}\$}\xi=\tilde{c}^{S^{\$},V^{\$}\$}.$$
In view of Equations \eqref{e:tilcsv} and \eqref{e:tilcs}, this yields
$$ \begin{pmatrix} \til{a}^{\top}\tilde{c}^{S\star}\til{a} & \til{a}^{\top} \tilde{c}^{S\star}R^{\top} \\ R \tilde{c}^{S\star}\til{a} & R \tilde{c}^{S\star}R^{\top} \end{pmatrix} \xi = \begin{pmatrix} \til{a}^{\top} \\ R \end{pmatrix} \left(\tilde{c}^{S,V\star}+\tilde{c}^{S\star}\til{a}V_{-}\right),$$
or equivalently, decomposed into the first and last $d$ components,
\begin{equation}\label{e:1}
\til{a}^{\top}\tilde{c}^{S\star}(\til{a},R^{\top})\xi=\til{a}^{\top}(\tilde{c}^{S,V\star}+\tilde{c}^{S\star}\til{a}V_{-})
\end{equation}
and
\begin{equation}\label{e:2}
R\tilde{c}^{S\star}(\til{a},R^{\top})\xi=R(\tilde{c}^{S,V\star}+\tilde{c}^{S\star}\til{a}V_{-}).
\end{equation}
By multiplying both sides of \eqref{e:1} with $S_{-}$ from the left and subtracting the result from \eqref{e:2}, this leads to
\begin{equation}\label{e:3}
\tilde{c}^{S\star}(\til{a},R^{\top})\xi=\tilde{c}^{S,V\star}+\tilde{c}^{S\star}\til{a}V_{-},
\end{equation}
since $R-S_{-}\til{a}^{\top}=E_d$. By Theorem \ref{s:KS},
\begin{equation*}
\phi'=(\til{a},R^{\top})\xi-(\pi^0+\phi' \mal S_{-})\til{a}
\end{equation*}
defines a marginal utility-based hedging strategy. Let
\begin{equation*}
\psi':=\phi'-((\tilde{c}^{S\star})^{-1}\tilde{c}^{S,V\star}-(\pi^0+\phi'\mal S_{-} -V_{-})\til{a})=(\til{a},R^{\top})\xi-(\tilde{c}^{S\star})^{-1}\tilde{c}^{S,V\star}-V_{-}\til{a}.
\end{equation*}
Then it follows from the definition of $\psi'$ and  \eqref{e:3} that
\begin{equation*}
\tilde{c}^{S\star}\psi' = \tilde{c}^{S,V\star}+\tilde{c}^{S\star}V_{-}\til{a}-\tilde{c}^{S,V\star}-\tilde{c}^{S\star}V_{-}\til{a}= 0,
\end{equation*}
because $\tilde{c}^{S\star}(\tilde{c}^{S\star})^{-1}\tilde{c}^{S,V\star}=\tilde{c}^{S,V\star}$ by \cite[Theorem 9.1.6]{albert.72}. In particular, $(\psi')^{\top}\tilde{c}^{S\star}\psi'=0$. Since $L^{\$}/L^{\$}_0=\E(K^{\$})>0$ and hence $\Delta K^{\$}>-1$ by \cite[I.4.61]{js.03}, this implies 
\begin{equation}\label{e:null}
(\psi')^{\top}\tilde{c}^{S}\psi'=0.
\end{equation}
For $n \in \nn$ define the predictable sets $D_n:=\{|\psi'|\leq n\}$. By Proposition \ref{p:Int} and \eqref{e:null}, we have $\tilde{c}^{\psi' 1_{D_n} \mal S}=0$ and hence $c^{\psi' 1_{D_n} \mal S}=0$ and $F^{\psi' 1_{D_n} \mal S}=0$. Together with Proposition \ref{p:Measure}, this implies that the local characteristics of $\psi' 1_{D_n} \mal S$ under the equivalent local martingale measure $Q$ from Assumption \ref{a:1} vanish by \cite[Lemma 3.1]{kallsen.03}. Hence $\psi' 1_{D_n} \mal S = 0$ and it follows from \cite[Lemma 2.2]{kallsen.03} that $\psi' \in L(S)$ with $\psi' \mal S=0$. Taking into account the definition of $\psi'$, this shows 
$$ \phi'\mal S= ((\tilde{c}^{S\star})^{-1}\tilde{c}^{S,V\star}-(\pi^0-V_{-})\til{a})\mal S-(\phi'\mal S_{-})\mal (\til{a}\mal S),$$
i.e., $\phi' \mal S$ solves the feedback equation 
\begin{equation}\label{e:feedback}
G = ((\tilde{c}^{S\star})^{-1}\tilde{c}^{S,V\star}-(\pi^0-V_{-})\til{a})\mal S - G_{-} \mal (\til{a} \mal S).
\end{equation}
Since $\psi' \in L(S)$ and $L(S)$ is a vector space, it follows that $(\tilde{c}^{S\star})^{-1}\tilde{c}^{S,V\star} \in L(S)$, too. As in the proof of \cite[Lemma 4.9]{cerny.kallsen.05}, this in turn yields that $\varphi'$ is well-defined and in $L(S)$. Evidently, $\varphi' \mal S$ also solves \eqref{e:feedback} and, since the solution is unique by \cite[(6.8)]{jacod.79}, we obtain $\varphi' \mal S=\phi'\mal S$. Therefore $\varphi'$ is a marginal utility-based hedging strategy.

We now turn to the risk premium $\pi'$. First notice that by \cite[Theorem 9.1.6]{albert.72},
\begin{gather*}
C^\$ :=\tilde{c}^{V^{\$}\$}-(\tilde{c}^{S^{\$},V^{\$}\$})^{\top}\xi=\tilde{c}^{V^\$ \$}-(\tilde{c}^{S^\$,V^\$ \$})^{\top}(\tilde{c}^{S^\$ \$})^{-1}\tilde{c}^{S^\$,V^\$ \$} \geq 0,\\
C^{\eur} := \tilde{c}^{V\star}-(\tilde{c}^{S,V\star})^{\top}(\tilde{c}^{S\star})^{-1}\tilde{c}^{S,V\star} \geq 0.
\end{gather*}
Hence $C^\$\mal A$ is an increasing predictable process and, by Lemmas \ref{l:Char} and \ref{l:measurechange},
\begin{align*}
E_{Q^\$}((N^{\$}_T)^2)&=E_{Q^{\$}}(C^{\$}\mal A_T)\\
&=\frac{C_1}{C_0}E_{P^{\eur}}\left(L^{\$}_{-}\E(-\til{a}\mal S)_{-}^2 C^{\$} \mal A_T\right)\\
&=\frac{C_1}{C_0}E_{P^{\eur}}\left(L^{\$}_{-}\E(-\til{a}\mal S)_{-}^2\mal (\langle V^{\$},V^{\$}\rangle_T^{Q^{\$}}-\langle V^{\$}, \xi \mal S^{\$}\rangle_T^{Q^{\$}})\right).
\end{align*}
Since we have shown $\phi'\mal S =\varphi' \mal S$ above, \cite[Proposition 2.1]{kallsen.goll.99a} and the proof of Theorem \ref{s:KS} yield $\xi \mal S^{\$}=(\varphi'^{0},\varphi')\mal S^{\$}$ for $\varphi'^{0}:=\pi^0+\varphi'\mal S-\varphi'S$. Hence 
\begin{align*}
E_{Q^\$}((N^{\$}_T)^2)&= \frac{C_1}{C_0}E_{P^{\eur}}\left(L^{\$}_{-}\E(-\til{a}\mal S)_{-}^2 \mal\left(\langle V^{\$},V^{\$}\rangle_T^{Q^{\$}}-\langle V^{\$},(\varphi'^{0},\varphi') \mal S^{\$} \rangle_T^{Q^{\$}}\right)\right)\\
                     &= \frac{C_1}{C_0}E_{P^{\eur}}\left(L^{\$}_{-}\E(-\til{a}\mal S)_{-}^2\left(\tilde{c}^{V^{\$}\$}-(\tilde{c}^{S^{\$},V^{\$}\$})^{\top}(\varphi'^{0},\varphi')\right)\mal A_T\right).
\end{align*}
After inserting $\tilde{c}^{V^\$ \$}$, $\tilde{c}^{S^\$,V^\$ \$}$ from \eqref{e:tilcv} resp.\ \eqref{e:tilcsv} and the definition of $(\varphi'^{0},\varphi')$, this leads to
\begin{equation}
E_{Q^\$}((N^{\$}_T)^2)=\frac{C_1}{C_0} E_{P^{\eur}}\left(\left(1+\Delta A^{K^\$}\right)L^{\$}_{-}C^{\eur}\mal A_T\right).\label{e:expectation}
\end{equation}
Now notice that the definition of the stochastic exponential and \cite[I.4.36]{js.03} imply
$$L^{\$}=\left(1+\Delta A^{K^\$}+\Delta M^{K^\$}\right)L^{\$}_{-}.$$
By \cite[I.4.49]{js.03} the process $ \Delta M^{K^\$} \mal (L^{\$}_{-}C^{\eur}\mal A)$ is a local martingale. If $(T_n)_{n \in \nn}$ denotes a localizing sequence, this yields
\begin{align*}
E_{P^{\eur}}(L^{\$}C^{\eur} \mal A_{T \wedge T_n}) &= E_{P^{\eur}}\left(\left(1+\Delta A^{K^\$}+\Delta M^{K^\$}\right)L^{\$}_{-}C^{\eur} \mal A_{T \wedge T_n}\right)\\
&= E_{P^{\eur}}\left(\left(1+\Delta A^{K^\$}\right)L^{\$}_{-}C^{\eur} \mal A_{T \wedge T_n}\right),
\end{align*}
and hence 
\begin{equation*}
E_{P^{\eur}}(L^{\$}C^{\eur} \mal A_{T}) =E_{P^{\eur}}\left(\left(1+\Delta A^{K^\$}\right)L^{\$}_{-}C^{\eur} \mal A_{T}\right)
\end{equation*}
by monotone convergence. Combining this with \eqref{e:expectation}, we obtain
\begin{equation*}
E_{Q^\$}((N_T^\$)^2)=\frac{C_1}{C_0}E_{P^{\eur}}(\left(\tilde{c}^{V\star}-(\tilde{c}^{S,V\star})^{\top}(\tilde{c}^{S\star})^{-1}\tilde{c}^{S,V\star}\right)L^{\$} \mal A_{T}).
\end{equation*}
In view of Theorem \ref{s:KS}, this completes the proof.\ep\\

\begin{bemm}
\begin{enumerate}
\item The arguments used to show $\varphi' \mal S=\phi' \mal S$ in the proof of Theorem \ref{s:mainresult} also yield that one obtains a marginal utility-based hedging strategy if the \emph{pure hedge coefficient} $(\tilde{c}^{S\star})^{-1}\tilde{c}^{S,V\star}$ is replaced by any other solution $\zeta$ of $\tilde{c}^{S\star}\zeta=\tilde{c}^{S,V\star}$. 
\item An inspection of the proof of Theorem \ref{s:mainresult} shows that the formulas for $\varphi'$ and $\pi'$ are independent of the specific semimartingale decomposition of $K^\$$ that is used. In particular, the not necessarily predictable term $1+\Delta A^{K^{\$}\eur}$ disappears in the formula for $\varphi'$ by \cite[Theorem 3.9]{albert.72}. If the semimartingale $K^{\$}$ is $P^{\eur}$-special, one can choose the \emph{canonical} decomposition \cite[II.2.38]{js.03}. By \cite[II.2.29]{js.03}, this yields 
\begin{equation*}\label{e:AK}
\Delta A^{K^{\$}}=\Delta A \int x F^{K^{\$}\eur}(dx).
\end{equation*}
If additionally  $K^{\$}$ has no fixed times of discontinuity,  \cite[II.2.9]{js.03} shows that $A$ can be chosen to be continuous, which implies $\Delta A^{K^{\$}}=0$.
\item For \emph{continuous} $S$, our feedback representation of $\varphi'$ coincides with \cite[Theorem 3]{kramkov.sirbu.06b} because the modified second characteristic is invariant with respect to equivalent changes of measure for continuous processes. 
\end{enumerate}
\end{bemm}

\section{Semi-explicit formulas in concrete models}\setcounter{equation}{0}

We now discuss how Theorem \ref{s:mainresult} can be applied in our concrete examples to yield numerically tractable representations of power utility-based prices and hedging strategies.

\subsection{Exponential L\'evy models}
For exponential L\'evy models, Theorem \ref{s:mainresult} indeed leads to a mean-variance hedging problem. Consequently, semi-explicit formulas for the objects of interest are provided by the results of Hubalek et al.\ \cite{hubalek.al.05} for mean-variance hedging in exponential L\'evy models. 

To this end, we fix a univariate exponential L\'evy model $S=S_0\E(X)>0$, with some non-monotone square-integrable L\'evy process $X$. Its L\'evy-Khintchine triplet relative to the truncation function $h(x)=x$ is denoted by $(b^X,c^X,F^X)$. Finally, we suppose throughout that the optimal fraction $\widehat{\eta}$ for the pure investment problem lies in the interior of the admissible fractions of wealth in stock $\scr{C}^0$, implying that all assumptions of Sections \ref{s:kramkov.sirbu} and \ref{s:alternative} are satisfied. 

\begin{bem}\label{bem:stoexp}
By \cite[Lemma A.8]{kallsen.goll.99a}, the stock price can also be written as the ordinary exponential $S=S_0\exp(\tilde{X})$ of the L\'evy process $\tilde{X}$ with L\'evy-Khintchine triplet
\begin{gather*}
b^{\tilde{X}}=b^X-\frac{1}{2}c^{X}+\int(\log(1+x)-x)F^X(dx), \quad c^{\tilde{X}}=c^X,\\
F^{\tilde{X}}(G)=\int 1_G(\log(1+x))F^X(dx) \quad \forall G \in \scr{B},
\end{gather*}
relative to $h(x)=x$.
\end{bem}

Since the density process $(L^{\eur}/L^{\eur}_0)\E(\widehat{\eta}X)^{-1-p}$ of $P^{\eur}$ with respect to $P$ is an exponential  L\'evy process, Proposition \ref{p:Measure} shows that $X$ is also a L\'evy process under $P^{\eur}$ with L\'evy-Khintchine triplet $(b^{X,\eur},c^{X,\eur},F^{X,\eur})$ given by
$$\left(b^X-(1+p)\widehat{\eta}c^X-\int x(1-(1+\widehat{\eta}x)^{-1-p})F^X(dx),c^X,(1+\widehat{\eta}x)^{-1-p}F^X(dx)\right),$$
relative to $h(x)=x$. This truncation function can be used because $X$ is square-integrable under $P^{\eur}$ as well by \cite[Corollary 25.8]{sato.99} and the proof of Lemma \ref{lem:levybound}. Moreover, since the budget constraint $\scr{C}^0$ is ``not binding,'' the first-order condition \cite[Equation (6.3)]{nutz.10c} implies that the drift rate $b^{X,\eur}$ can also be written as
\begin{equation}\label{eq:beur}
b^{X,\eur}=-\widehat{\eta}\left(c^X+\int \frac{x^2}{(1+\widehat{\eta}x)^{1+p}} F^X(dx)\right)=-\widehat{\eta}\tilde{c}^{X,\eur}.
\end{equation}

\begin{lemma}\label{lem:adm}
The optimal trading strategy  $\widehat{\varphi}=-v\til{a}\E(-\til{a}\mal S)$ in the pure investment problem is admissible in the sense of  \cite[Corollary 2.5]{cerny.kallsen.05}.
\end{lemma}

\bpf In view of \cite[Proposition 13]{schweizer.94}, the mean-variance optimal hedge for the constant claim $H=v$ is $-v\til{\lambda}\E(-\til{\lambda}\mal S)$ for $\til{\lambda}=b^{X,\eur}/(S_{-}\tilde{c}^{X,\eur})$. Hence $\til{a}=\til{\lambda}$ by \eqref{eq:beur}; in particular, $\widehat{\varphi}$ is admissible in the sense of Schweizer \cite{schweizer.94} and therefore in the sense of \v{C}ern\'y and Kallsen \cite{cerny.kallsen.05} as well by \cite[Corollary 2.9]{cerny.kallsen.05}.
\ep \\

Together with the discussion at the end of Section 4, Lemma \ref{lem:adm} immediately yields 

\begin{cor}\label{c:hedge}
Let $H$ be a contingent claim satisfying Assumption \ref{a:5}. Then  the marginal utility-based price $\pi^0$, the marginal utility-based hedging strategy $\varphi'$, and the risk premium $\pi'$ from Theorem \ref{s:mainresult} coincide with the mean-variance optimal initial capital, the mean-variance optimal hedge and the $p\exp((a^{\eur}-a)T)/2v$-fold of the minimal expected squared hedging error $\epsilon_{\eur}^2$ for $H$ under $P^{\eur}$.
\end{cor} 

Corollary \ref{c:hedge} implies that -- in first-order approximation --  power utility-based hedging corresponds to mean-variance hedging, but for a L\'evy process with different drift and jump measure. If $\eta=0$, which is equivalent to $S$ being a martingale under the physical measure $P$, then $P^{\eur}=P$ and no adjustment is necessary. If $\widehat{\eta}>0$ in the economically most relevant case of a positive drift, the stock price process is a $P$-submartingale, but turns into a supermartingale under $P^{\eur}$. Moreover, negative jumps become more likely and positive jumps less likely, such that a negative skewness is amplified when passing from $P$ to $P^{\eur}$. The magnitude of these effects depends on the investor's risk aversion $p$. Note that as the latter becomes large, the $P^{\eur}$ dynamics of the return process converge to those under the minimal entropy martingale measure (cf., e.g., \cite{fujiwara.miyahara.03}). Hence, as risk aversion becomes large, asymptotic power utility-based pricing and hedging approaches its counterpart for exponential utility. 

The above considerations apply to \emph{any} contingent claim satisfying Assumption \ref{a:5}, i.e., which can be superhedged with respect to the numeraire given by the optimal wealth process in the pure investment problem. To obtain numerically tractable formulas, one has to make additional assumptions. For example, semi-explicit solutions to the mean-variance hedging problem for exponential L\'evy models have been obtained in \cite{hubalek.al.05} using the \emph{Laplace transform approach} put forward in \cite{raible.00}. The key assumption for this approach is the existence of an integral representation of the payoff function in the following sense.

\begin{assumption}\label{a:intrep}
Suppose $H=f(S_T)$ for a function $f: (0, \infty) \mapsto \rr$ such that
$$ f(s)=\int_{R-i\infty}^{R+i\infty} l(z) s^z dz, \quad s \in (0,\infty),$$
for $l:\mathbb{C} \to \mathbb{C}$ such that the integral exists for all $s \in (0,\infty)$ and $R \in \rr$ such that $E(S_T^R)<\infty$.
\end{assumption} 

Most European options admit a representation of this kind, see, e.g., \cite[Section 4]{hubalek.al.05}. 

\begin{bsp}\label{bsp:intrepcall}
For a European call option with strike $K > 0$ we have $H=(S_T-K)^+$ and, for $s>0$ and $R>1$,
$$ (s-K)^+=\frac{1}{2\pi i}\int_{R-i\infty}^{R+i\infty} \frac{K^{1-z}}{z(z-1^)}s^z dz.$$
\end{bsp}

By evaluating the formulas of Hubalek et al.\ \cite{hubalek.al.05} under $P^{\eur}$, we obtain the following semi-explicit representations. They are expressed in terms of the L\'evy exponent $\psi^{\tilde{X} \eur}$ of the log-price $\tilde{X}$ under $P^{\eur}$.

\begin{satz}\label{t:hubalek}
For a contingent claim $H$ satisfying Assumptions \ref{a:5} and \ref{a:intrep}, the marginal utility-based price and a marginal utility-based hedging strategy are given by 
\begin{align*}
\pi(0)&=V_0,\\
\varphi'_t &= \xi_t -\left(V_0+\varphi' \mal S_{t-}-V_{t-}\right)\til{a},
\end{align*}
with 
\begin{gather*}
\Psi(z):=\psi^{\tilde{X}\eur}(z)-\psi^{\tilde{X}\eur}(1)\frac{\psi^{\tilde{X}\eur}(z+1)-\psi^{\tilde{X}\eur}(z)-\psi^{\tilde{X}\eur}(1)}{\psi^{\tilde{X}\eur}(2)-2\psi^{\tilde{X}\eur}(1)},\\
\widetilde{a}:=\frac{1}{S_{t-}}\frac{\psi^{\tilde{X}\eur}(1)}{\psi^{\tilde{X}\eur}(2)-2\psi^{\tilde{X}\eur}(1)},\\
V_t:=\int_{R-i\infty}^{R+i\infty} S_t^z e^{\Psi(z)(T-t)} l(z)dz,\\
\xi_t:= \int_{R-i\infty}^{R+i\infty} S_{t-}^{z-1} \frac{\psi^{\tilde{X}\eur}(z+1)-\psi^{\tilde{X}\eur}(z)-\psi^{\tilde{X}\eur}(1)}{\psi^{\tilde{X}\eur}(2)-2\psi^{\tilde{X}\eur}(1)}  e^{\Psi(z)(T-t)} l(z) dz.
\end{gather*}
Moreover, the corresponding risk premium $\pi'$ for $H$ can be written as
\begin{align*}
\pi'= \frac{p \exp((a^{\eur}-a)T)}{2v}\int_{R-i\infty}^{R+i\infty} \int_{R-i\infty}^{R+i\infty} J(z_1,z_1) l(z_1) l(z_2) dz_1 dz_2,
\end{align*}
for $a$ and $a^{\eur}$ as in Examples \ref{bsp:levyinvest} and \ref{bsp:levyleur}, respectively, and
\begin{align*}
k(z_1,z_2)&:=\Psi(z_1)+\Psi(z_2)-\frac{\psi^{\tilde{X}\eur}(1)^2}{\psi^{\tilde{X}\eur}(2)-2\psi^{\tilde{X}\eur}(1)},\\
j(z_1,z_2) &:= \psi^{\tilde{X}\eur}(z_1+z_2)-\psi^{\tilde{X}\eur}(z_1)-\psi^{\tilde{X}\eur}(z_2)\\
\quad \quad \quad \quad &-\frac{(\psi^{\tilde{X}\eur}(z_1+1)-\psi^{\tilde{X}\eur}(z_1)-\psi^{\tilde{X}\eur}(1))(\psi^{\tilde{X}\eur}(z_2+1)-\psi^{\tilde{X}\eur}(z_2)-\psi^{\tilde{X}\eur}(1))}{\psi^{\tilde{X}\eur}(2)-2\psi^{\tilde{X}\eur}(1)},\\
J(z_1,z_2)&:=\begin{cases} S_0^{z_1+z_2} j(z_1,z_2) \displaystyle \frac{e^{k(z_1,z_2)T}-e^{\psi^{\tilde{X}\eur}(z_1+z_2)T}}{k(z_1,z_2)-\psi^{\tilde{X}\eur}(z_1+z_2)} &\mbox{if}\ k(z_1,z_2) \neq \psi^{\tilde{X}\eur}(z_1+z_2),\\ S_0^{z_1+z_2}j(z_1,z_2)T e^{\psi^{\tilde{X}\eur}(z_1,z_2)T} &\mbox{if}\ k(z_1,z_2)=\psi^{\tilde{X}\eur}(z_1,z_2). \end{cases}
\end{align*}
\end{satz}

\bpf See \cite[Theorems 3.1 and 3.2]{hubalek.al.05}. \ep

\subsection{BNS model}\label{subsec:bns}

We now turn to the application of Theorem \ref{s:mainresult} to the BNS model with stochastic volatility. Throughout, we assume that the conditions of Examples \ref{bsp:wellposed} and \ref{bsp:levyleur} are satisfied, i.e., either $p \geq 2$ or sufficiently large exponential moments of the subordinator $Z$ driving the variance process $y$ exist. In the first case, we also suppose $Z$ is integrable. By Proposition \ref{p:Measure}, the $P^{\eur}$-dynamics of the variance process $y$ and the return process $X$ are given by
\begin{align*}
dy_t &= -\lambda dt + dZ^{\eur}_t,\\
dX_t &= (-\mu/p)y_tdt+\sqrt{y_t}dW_t.
\end{align*}
Here $\mu$ and $\lambda$ are the constant drift and mean reversion rates of the BNS model under $P$, $W$ is a standard Brownian motion (under both $P$ and $P^{\eur}$), and $Z^{\eur}$ is an \emph{inhomogeneous} $P^{\eur}$-L\'evy process with characteristics 
$$\left(b^Z+\int_0^\infty z(e^{\alpha_1^{\eur}(t)z}-1)F^Z(dz),0,e^{\alpha_1^{\eur}(t)z}F^Z(dz)\right)$$
relative to the truncation function $h(z)=z$. Hence $(y,X)$ is an \emph{inhomogeneous} BNS model under $P^{\eur}$. Note that as for exponential L\'evy models, the drift rate of the return process changes its sign when moving from $\mu$ (under $P$) to $-\mu/p$ (under $P^{\eur}$). The effect on the volatility process $y$ depends on the sign of $\alpha_1^{\eur}$, which is positive for $p<2$ and negative for $p>2$. If $p<2$, i.e., for less risk-averse investors, the mean of $Z^{\eur}$ (i.e., the average size of the positive volatility jumps) increases because jumps (in particular, large ones) become more likely under $P^{\eur}$. For more risk averse investors with $p>2$, the frequency of jumps is decreased under $P^{\eur}$, which also leads to a decrease in the average value of volatility. Since $\alpha^{\eur}_1(t)$ decreases resp.\ increases to $0$ as $t \to T$ for $p<2$ resp.\ $p>2$, the deviation from the $P$-dynamics of $Z$ is largest at the initial time $t=0$ and tends to zero as $t \to T$. Finally, as the investor's risk aversion becomes large, the $P^{\eur}$ dynamics of $(y,X)$ again tend to their counterparts under the minimal entropy martingale measure corresponding to exponential utility, which was determined in \cite{benth.meyerbrandis.05}.

With the $P^{\eur}$-dynamics of $S$ at hand, we can now provide a sufficient condition for the validity of Assumption \ref{a:6} in the BNS model. More specifically, $S$ is square-integrable under $P^{\eur}$ by \cite[Theorem 5.1]{kallsen.muhlekarbe.10b} provided that
$$ \int_1^\infty \exp\left(\frac{1-e^{-\lambda T}}{\lambda} \left(\frac{(1+p)(2-p)}{2p^2}\mu^2+2-\frac{\mu}{p}\right)z\right)F^Z(dz)<\infty.$$
If, in addition, the conditions of Example \ref{bsp:levyleur} are satisfied, Assumption \ref{a:6} holds.

We now turn to the computation of semi-explicit representations for the marginal utility-based price $\pi^0$ (cf.\ Remark \ref{rem:qopt}) as well as the utility-based hedge $\varphi'$ and the risk premium $\pi'$ from Theorem \ref{s:mainresult} for claims admitting an integral representation as in Assumption \ref{a:intrep}. The (inhomogeneous) BNS model is studied from the point of view of mean-variance hedging in \cite{kallsen.vierthauer.09}. As noted in the introduction, the formulas in Theorem \ref{s:mainresult} formally agree with such a problem under the appropriate probability measure $P^{\eur}$. Therefore the calculations in \cite{kallsen.vierthauer.09} can be adapted to the present situation. In that paper, admissibility of the candidate solution $\til{a}$ to the pure investment problem under quadratic utility is not shown. Nevertheless, the results from \cite{kallsen.vierthauer.09} can be applied here because $\tilde{a}$ does not have to be admissible for the application of Theorem \ref{s:mainresult}. Put differently, the calculations in \cite{kallsen.vierthauer.09} can be used without explicitly referring to the quadratic hedging problem studied there. Below, we outline the necessary steps. This sketch could be turned into a rigorous proof, similarly as in \cite[Theorems 4.1 and 4.2]{kallsen.pauwels.09a}. 

The \emph{first step} is to determine the mean value process $V=E_{Q_0}(H|\scr{F}_t)$. Since the density process $L\E(\widehat{\eta}X)^{-p}$ of $Q_0$ with respect to $P$ is the exponential of an inhomogeneous affine process (cf.\ \cite{filipovic.05,kallsen.muhlekarbe.10b} for more details), Proposition \ref{p:Measure} shows that $(y,X)$ is also an inhomogeneous affine process under $Q_0$. Using the integral representation for $H$, Fubini's theorem, and the affine transform formula for $(y,X)$ (compare \cite{filipovic.05,kallsen.muhlekarbe.10b}) then leads to 
\begin{equation}\label{eq:mv}
V_t            = \int_{R-i\infty}^{R+i\infty} S_{t}^z \exp\left(\Psi^0(t,T,z)+\Psi^1(t,T,z)y_{t}\right)l(z) dz,
\end{equation}
with 
\begin{align*}
\Psi^1(t,T,z)  &= \frac{(1-z)z}{2\lambda}(e^{-\lambda(T-t)}-1),\\
\Psi^0(t,T,z)  &= \int_t^T \left( \psi^Z(\alpha_1(s)+\Psi^1(s,T,z))-\psi^Z(\alpha_1(s))\right)ds.
\end{align*}

In the \emph{second step}, we turn to the marginal utility-based hedging strategy $\varphi'$. The representation \eqref{eq:mv} for $V$ and the bilinearity of the predictable quadratic variation yields integral representations for the modified second $P^{\eur}$-characteristics of $(S,V)$, too, where the integrands can be computed using Proposition \ref{p:Ito} (cf.\ the proof of \cite[Theorem 3.3]{kallsen.vierthauer.09} for more details). Plugging these in Theorem \ref{s:mainresult} gives 
$$\varphi'=\xi_t-(V_0+\varphi' \mal S_{t-} -V_{t-})\til{a}_t,$$
with
$$\xi_t          = \int_{R-i\infty}^{R+i\infty} z S_{t}^{z-1} \exp\left(\Psi^0(t,T,z)+\Psi^1(t,T,z)y_{t-}\right) l(z) dz,$$
for $\Psi_0,\Psi_1$ as above. 

\begin{bem}\label{rem:bnssquare}
Provided that differentiation and integration can be interchanged, the \emph{pure hedge coefficient} $\xi_t$ in the BNS model is given by the derivative of $V_t$ with respect to $S_t$. Hence the marginal utility-based hedging strategy is given as the sum of the \emph{delta hedge} with respect to the marginal utility-based option price and a feedback term. This is a generic result in affine models with continuous asset prices and uncorrelated volatility processes, compare \cite{kallsen.vierthauer.09}.
\end{bem}

Finally, in a \emph{third step}, it remains to consider the risk premium $\pi'$ in Theorem \ref{s:mainresult}. Plugging in the expression for $L$ and $L^{\eur}$, we find
$$\frac{pC_1}{2vC_0}=\frac{pL^{\eur}_0}{2vL_0}=\frac{p}{2v} \exp\left(\int_0^T (\psi^Z(\alpha_1^{\eur}(t))-\psi^Z(\alpha_1(t)))dt+\frac{\mu^2(1-e^{-\lambda T})}{p^2 \lambda}y_0\right).$$
Hence it remains to compute the expectation in the formula for the risk premium $\pi'$.  Here, \eqref{eq:mv} again leads to integral representations for $\tilde{c}^{S\star},\tilde{c}^{S,V\star},\tilde{c}^{V\star}$. The product of $L^{\$}$ and the integrand once more turns out to be the exponential of an inhomogeneous affine process. Its expectation can therefore again be computed using the affine transform formula for $(y,X)$ (cf.\ the proof of \cite[Theorem 3.4]{kallsen.vierthauer.09} for more details). This leads to
\begin{equation*}
\pi'= \frac{p C_1}{2v C_0}\int_0^T \int_{R-i\infty}^{R+i\infty} \int_{R-i\infty}^{R+i\infty} J(t,z_1,z_2) l(z_1)l(z_2) dz_1 dz_2 dt,
\end{equation*}
for
\begin{align*}
\psi^{Z\eur}(t,u)&=\psi^Z(u+\alpha_1^{\eur}(t))-\psi^Z(\alpha_1^{\eur}(t)),\\
j(t,z_1,z_2)               =&\psi^{Z\eur}(t,\alpha_1^{\$}(t)+\Psi^1(t,T,z_1)+\Psi^1(t,T,z_2))+\psi^{Z\eur}(t,\alpha_1^{\$}(t))\\
&-\psi^{Z\eur}(t,\alpha_1^{\$}(t)+\Psi^1(t,T,z_1))-\psi^{Z\eur}(t,\alpha_1^{\$}(t)+\Psi^1(t,T,z_2)),\\
g(z_1,z_1)               =&\frac{2\mu+p}{2p}(z_1+z_2)-\frac{1}{2}(z_1+z_2)^2,
\end{align*}
and 
\begin{align*}                              
\Upsilon^1(s,t,T,z_1,z_2)  =&(\alpha_1^{\$}(t)+\Psi^1(t,T,z_1)+\Psi^1(t,T,z_2))e^{\lambda(s-t)}+g(z_1,z_2)\frac{e^{\lambda(s-t)}-1}{\lambda},\\
\Upsilon^0(s,t,T,z_1,z_2)    =& \int_s^t \psi^{Z\eur}(r,\Upsilon^1(r,t,T,z_1,z_2))dr,\\
J(t,z_1,z_2)=&S_0^{z_1+z_2} j(t,z_1,z_2) \exp\left(\Upsilon^0(0,t,T,z_1,z_2)+\Upsilon^1(0,t,T,z_1,z_2)y_0\right)\\
           &\times \exp\left(\int_t^T \psi^{Z\eur}(s,\alpha_1^{\$}(s))ds+\Psi^0(t,T,z_1)+\Psi^0(t,T,z_2)\right).
\end{align*}
If the volatility process $y$ is chosen to be a Gamma-OU process, all expressions involving integrals of the characteristic exponent $\psi^{Z\eur}(t,u)=\psi^Z(u+\alpha_1^{\eur}(t))-\psi^Z(\alpha_1^{\eur}(t))$ can be computed in closed form as well. More specifically, let $y$ be a Gamma-OU process with mean reversion rate $\lambda>0$ and stationary $\Gamma(a,b)$-distribution and let
$$m(s):=c_1\left(e^{-\lambda(\til{t}-s)}-1\right)+c_2e^{-\lambda(\til{t}-s)}+c_3, \quad \til{t} \in [0,T],$$
for constants $c_1,c_2,c_3 \in \mathbb{C}$. Then if $m(s) \neq b$, $s \in [t,T]$ we have
$$ \int_{t_1}^{t_2} \psi^Z(m(s))ds = 
\begin{cases} 
\frac{-a}{b+c_1-c_3}\left(\lambda (t_2-t_1)(c_1-c_3)-b\log\left(\frac{-b+m(t_1)}{-b+m(t_2)}\right)\right) & b \neq c_3-c_1, \\
-\lambda a (t_2-t_1) + \frac{a b}{c_1+c_2} \left( e^{  \lambda(\til{t}-t_2) } - e^{ \lambda(\til{t}-t_1) } \right) & b = c_3-c_1
\end{cases}
$$
for $0 \leq t_1 \leq t_2 \leq T$ and where $\log$ denotes the \emph{distinguished logarithm} in the sense of \cite[Lemma 7.6]{sato.99}. This follows by inserting the L\'evy exponent $\psi^Z(u)=\frac{\lambda a u}{b-u}$, which is analytic on $\mathbb{C} \backslash \{b\}$, and integration using decomposition into partial fractions.

\section{Numerical illustration}

\begin{figure}[htbp] 
\centering
\includegraphics[width=12cm]{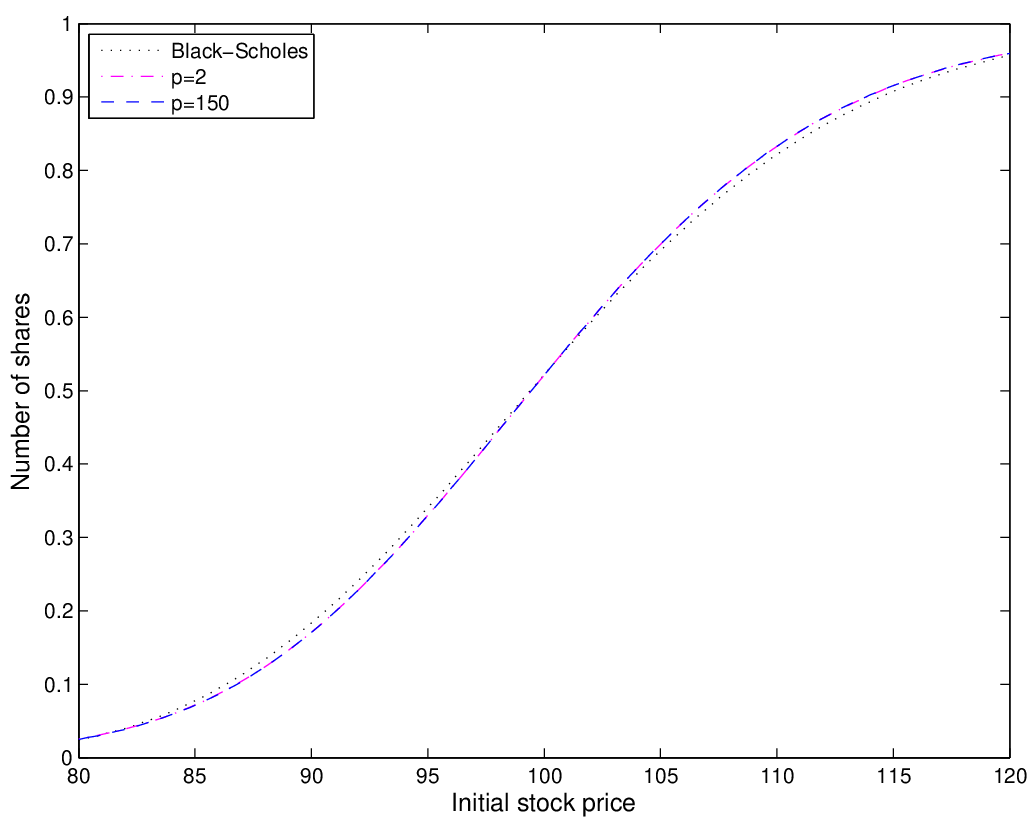}
\caption{Initial Black-Scholes hedge and initial utility-based BNS-hedges for $p=2$, $p=150$ and a European call with strike $K=100$ and maturity $T=0.25$.}\label{strategy}
\end{figure}

Mean-variance hedging for the BNS Gamma-OU stochastic volatility model is considered in \cite{kallsen.vierthauer.09}. Since the formulas in the previous section are of the same form, the numerical algorithm applied in \cite{kallsen.vierthauer.09} can also be used to explore this model from the point of view of utility-based pricing and hedging. Exponential L\'evy processes could be treated analogously (compare \cite{hubalek.al.05}). Since the corresponding results are very similar, we omit them here.

\begin{figure}[thbp] 
\centering
\includegraphics[width=12cm]{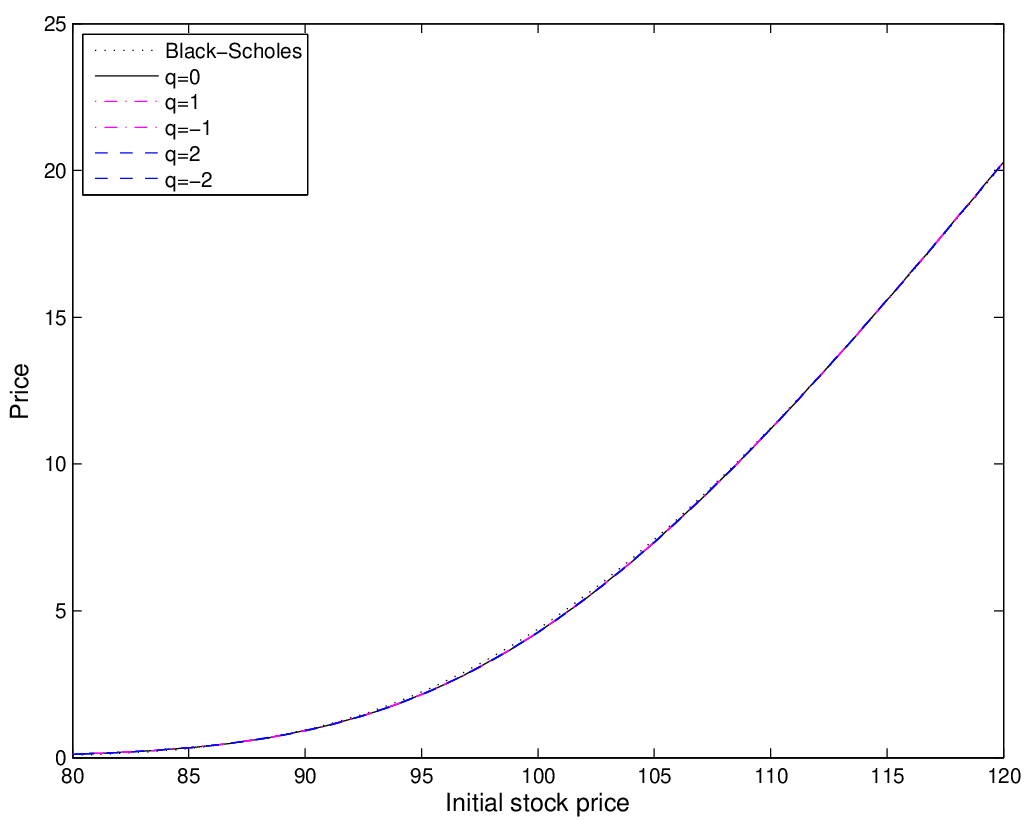}
\caption{Black-Scholes price and approximate indifference price $\pi(0)+q\pi'$ in the BNS model for $p=2$ and a European call with strike $K=100$ and maturity $T=0.25$.}\label{prices2}
\end{figure}

\begin{figure}[htbp] 
\centering
\includegraphics[width=12cm]{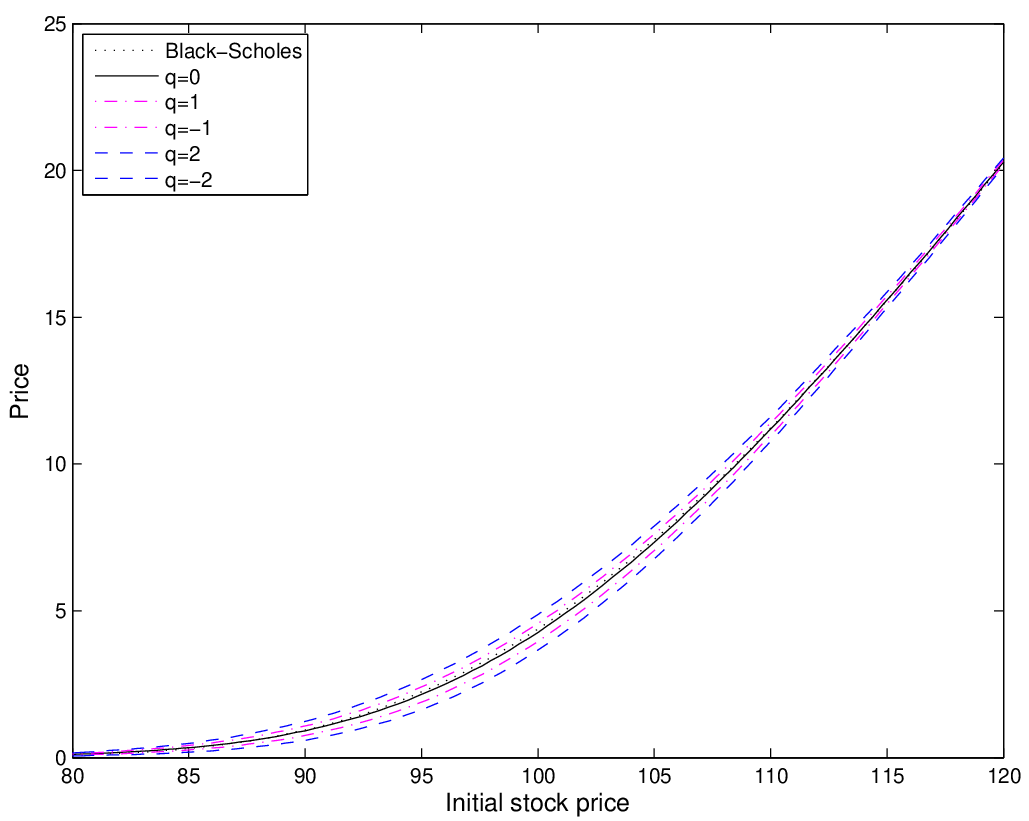}
\caption{Black-Scholes price and approximate indifference price $\pi(0)+q\pi'$ in the BNS model for $p=150$ and a European call with strike $K=100$ and maturity $T=0.25$.}\label{prices150}
\end{figure}

As a concrete specification, we consider the discounted BNS-Gamma-OU model with parameters as estimated in \cite{kallsen.muhlekarbe.10c} from a DAX time series, i.e.,
$$\mu=1.404, \quad \lambda=2.54, \quad a=0.848, \quad b=17.5.$$ 
We let $y_0=0.0485$ and put $v=241$, which implies that indifference prices and utility-based hedging strategies exist for $S_0 \in [80,120]$ and $q \in [-2,2]$. By our above results, first-order approximations of the utility-indifference price and the utility-based hedging strategy exist for $p=0.5, \ldots,150$ by Lemma \ref{l:exindiffprice} resp.\ Theorem \ref{t:exindiffhedge}. Moreover, Assumptions \ref{a:5} and \ref{a:intrep} hold for European call-options by Example \ref{bsp:intrepcall}. The formulas from Section \ref{subsec:bns} can now be evaluated using numerical quadrature, where we use  $R=1.2$. 

The initial hedges for $p=2$ and $p=150$ in Figure \ref{strategy} below cannot be distinguished by eye. Indeed, the maximal relative difference between the two strategies is $0.4\%$ for $80 \leq S_0 \leq 120$, which implies that the utility-based hedging strategy is virtually independent of the investor's risk aversion. Moreover, both strategies are quite close to the Black-Scholes hedging strategy, the maximal relative difference being about 8.9\%.

We now turn to utility-based pricing. First, note that in our specification the marginal utility-based price $\pi^0$ barely depends on the investor's risk aversion, and is almost indistinguishable from its Black-Scholes counterpart. For a relative risk aversion of $p=2$, the effect of the first-order risk adjustment is also very small (cf.\ Figure \ref{prices2}). This resembles similar findings of \cite{henderson.02, henderson.hobson.02} on utility-based pricing and hedging for basis risk.

\begin{figure}[hbtp] 
\centering
\includegraphics[width=12cm]{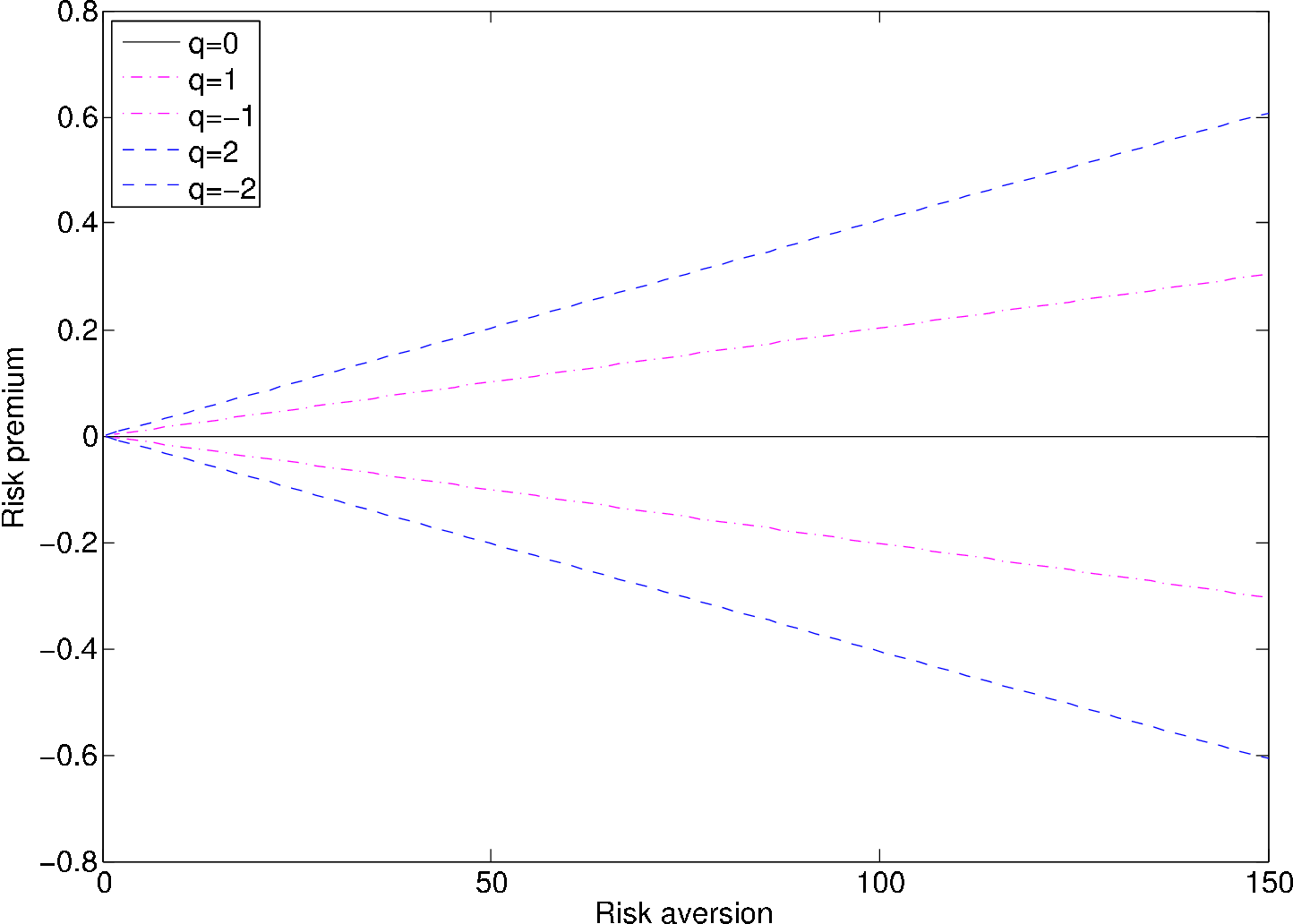}
\caption{Risk premia $q\pi'$ for $q$ at-the-money European calls with strike $K=100$ and maturity $T=0.25$ in the BNS model for risk aversions $p=0.5,\ldots,150$.}\label{RiskPremium1}
\end{figure}

In fact, much higher risk aversions as, e.g., $p=150$ in Figure \ref{prices150} are required to obtain a bid price below and an ask-price above the Black-Scholes price for one option as a result of the first-order risk adjustment. For evidence supporting such high levels of risk aversion, cf., e.g., \cite{janacek.04}. Finally, Figure \ref{RiskPremium1} depicts the dependence of the risk premium $\pi'$ on the investor's relative risk aversion $p$, which turns out to be almost linear. Note that since $\pi'$ is inversely proportional to the initial endowment $v$, this also implies that $\pi'$ is virtually linear in the investor's absolute risk aversion $p/v$, which holds exactly for exponential utility (cf.\ \cite{mania.schweizer.05,becherer.06,kallsen.rheinlaender.08}).

\begin{appendix}
\section{Appendix}\label{appendix}\setcounter{equation}{0}
In this appendix we summarize some basic notions regarding semimartingale characteristics (cf.\ \cite{js.03} for more details). In addition, we state and prove an auxiliary result which is used in the proof of Theorem \ref{s:mainresult}.

To any $\rr^d$-valued semimartingale $X$ there is associated a triplet $(B,C,\nu)$ of {\em characteristics}\index{semimartingale characteristics}, where $B$ resp.\ $C$ denote $\rr^d$- resp.\ $\rr^{d\times d}$-valued  predictable processes and $\nu$ a random measure on $\rp\times\rr^d$ (cf.\  \cite[II.2.6]{js.03}). The first characteristic $B$ depends on a {\em truncation function} $h: \rr^d \to \rr^d$ such as $h(x)=x 1_{\{|x| \leq 1\}}$. Instead of the characteristics themselves, we typically use the following notion.

\begin{defi}\label{d:diffchar}
Let $X$ be an $\rr^d$-valued semimartingale with characteristics $(B,C,\nu)$ relative to some truncation function $h$ on $\rr^d$. In view of \cite[II.2.9]{js.03}, there exist a predictable process $A \in \apl$, an $\rr^d$-valued predictable process $b$, an $\rr^{d \times d}$-valued predictable process $c$ and a transition kernel $F$ from $(\Omega \times \rp,\scr{P})$ into $(\rr^d,\B^d)$ such that  
$$ B_t=b \mal A_t, \quad C_t=c \mal A_t, \quad \nu([0,t] \times G)=F(G) \mal A_t \quad \mbox{for } t \in [0,T],\ G \in \B^d,$$
where we implicitly assume that $(b,c,F)$ is a good version in the sense that the values of $c$ are non-negative symmetric matrices, $F_s(\{0\})=0$ and $\int(1\wedge|x|^2)F_s(dx)<\infty$. We call $(b,c,F,A)$ \textit{local characteristics} of $X$.
\end{defi}

If $(b,c,F,A)$ denote local characteristics of some semimartingale $X$, we write 
$$ \tilde{c}:= c+ \int x x^{\top} F(dx)$$
and call $\tilde{c}$ the \textit{modified second characteristic} of $X$ provided that the integral exists. This notion is motivated by the fact that $\left\langle X,X \right\rangle = \tilde{c} \mal A$ by \cite[I.4.52]{js.03} if the corresponding integral is finite. We write $(b^X,c^X,F^X,A)$ and $\tilde{c}^X$ for the differential characteristics and the modified second characteristic of a semimartingale $X$. Likewise, the joint local characteristics of two semimartingales $X$, $Y$ are denoted by 
$$ (b^{(X,Y)},c^{(X,Y)},F^{(X,Y)},A)=\left(\begin{pmatrix} b^X \\ b^Y\end{pmatrix}, \begin{pmatrix} c^X & c^{X,Y} \\ c^{Y,X} & c^Y\end{pmatrix}, F^{(X,Y)},A\right)$$
and 
$$ \tilde{c}^{(X,Y)}=\begin{pmatrix} \tilde{c}^X & \tilde{c}^{X,Y} \\ \tilde{c}^{Y,X} & \tilde{c}^Y\end{pmatrix},$$
if the modified second characteristic of $(X,Y)$ exists. The characteristics of a semimartingale $X$ under some other measure $Q^{\$}$ are denoted by $(b^{X\$},c^{X\$},F^{X\$},A)$. The following rules for the computation of characteristics are used repeatedly in the proofs of this paper.

\begin{prop}[$C^2$-function]\label{p:Ito}
Let X be an $\rr^d$-va\-lued semimartingale with local characteristics $(b^X,c^X,F^X,A)$. Suppose that $f:U\to\rr^n$ is twice continuously differentiable on some open subset $U\subset\rr^d$ such that $X$,$X_-$ are $U$-valued. Then the $\rr^n$-va\-lued semimartingale $f(X)$ has local characteristics $(b^{f(X)},c^{f(X)},F^{f(X)},A)$, where
\beaa
b^{f(X),i}_t &=& \sum_{k=1}^{d}\partial_kf^i(X_{t-})b_t^{X,k}+\frac{1}{2}\sum_{k,l=1}^d\partial_{kl}f^i(X_{t-})c_t^{X,kl} \\
						 & & {} + \int\left(\til{h}^i(f(X_{t-}+x)-f(X_{t-}))-\sum_{k=1}^d\partial_kf^i(X_{t-})h^k(x)\right)F^X_t(dx),
\eeaa
as well as 
\beaa						 
c^{f(X),ij}_t &=& \sum_{k,l=1}^d\partial_kf^i(X_{t-})c_t^{X,kl}\partial_lf^j(X_{t-}),\\
F^{f(X)}_t(G) &=& \int1_G(f(X_{t-}+x)-f(X_{t-}))F^X_t(dx)\quad\forall G\in \B^n \mbox{ with } 0\notin G.
\eeaa
Here, $\partial_k$ etc.\ denote partial derivatives and $\til{h}$ again the truncation function on $\rr^n$.
\end{prop}

\bpf This follows immediately from \cite[Corollary A.6]{kallsen.goll.99a}. \ep\\ 

\begin{prop}[Stochastic integration]\label{p:Int}
Let X be an $\rr^d$-valued semimartingale with local characteristics $(b^X,c^X,F^X,A)$ and H an $\rr^{n\times d}$-valued predictable process with $H^{j\cdot}\in L(X)$ for $ j=1,\dots,n$. Then local characteristics of the $\rr^n$-valued integral process $H\mal X:=( H^{j\cdot}\mal X)_{j=1,\dots,n}$ are given by $(b^{H \mal X},c^{H \mal X},F^{H \mal X},A)$, where
\beaa
b^{H \mal X}_t &=& H_tb^X_t+\int(\til{h}(H_tx)-H_th(x))F^X_t(dx), \\
c^{H \mal X}_t &=& H_tc^X_tH_t^\top,\\
F^{H \mal X}_t(G) &=& \int 1_G(H_tx)F^X_t(dx)\quad \forall G\in\B^n \mbox{ with } 0\notin G.
\eeaa
Here $\til{h}:\rr^n\to\rr^n$ denotes the truncation function which is used on $\rr^n$.
\end{prop}

\bpf \cite[Lemma 3]{kallsen.shiryaev.00}. \ep\\

Let $P^{\star} \stackrel{\mathrm{loc}}{\sim} P$ be a probability measure with density process $Z$. Local equivalence yields that $Z$ and $Z_{-}$ are strictly positive by \cite[I.2.27]{js.03}. Hence the \emph{stochastic logarithm} \index{logarithm!stochastic} $N:=\L(Z)=\frac{1}{Z_{-}}\mal Z$ is a well-defined semimartingale. For an $\rr^d$-valued semimartingale $X$ we now have the following result, which relates the local $P^{\star}$-characteristics of $(X,N)$ to the local characteristics of $(X,N)$ under $P$.

\begin{prop}[Equivalent change of measure]\label{p:Measure}
Local $P^{\star}$-characteristics of the process $(X,N)$ are given by $(b^{(X,N)\star},c^{(X,N)\star},F^{(X,N)\star},A)$, where
\begin{align*}
b^{(X,N)\star}&=b^{(X,N)}+c^{(X,N),N}+\int h(x)x_{d+1}F^{(X,N)}(dx),\\ 
c^{(X,N)\star}&=c^{(X,N)},\\
F^{(X,N)\star}&=\int 1_G(x) (1+x_{d+1})F^{(X,N)}(dx) \quad \forall G\in \B^{d+1} \mbox{ with } 0\notin G.
\end{align*}
\end{prop}

\bpf \cite[Lemma 5.1]{kallsen.03}. \ep\\

The following observation is needed in the proof of Theorem \ref{s:mainresult}.

\begin{lemma}\label{l:measurechange}
Let $Q  \stackrel{\mathrm{loc}}{\ll} P$ with density process $Z$. Then for any increasing, predictable process $A$ with $A_0=0$ we have 
$$ E_{Q}(A_T)=E_{P}(Z_{-} \mal A_T).$$
\end{lemma}

\bpf Since $Z$ is a $P$-martingale and $A$ is predictable and of finite variation, $A \mal Z$ is a local $P$-martingale by \cite[I.3.10 and I.4.34]{js.03}. If $(T_n)_{n \in \nn}$ denotes a localizing sequence, $A \mal Z_{T \wedge T_n}$ is a martingale starting at $0$. By \cite[III.3.4 and I.4.49]{js.03}, this implies
\begin{align*}
E_{Q}(A_{T \wedge T_n}) = E_P(Z_{T \wedge T_n} A_{T \wedge T_{n}})&= E_P(Z_{-} \mal A_{T \wedge T_n}+A \mal Z_{T \wedge T_n})= E_P(Z_{-} \mal A_{T \wedge T_n}).
\end{align*}
Hence monotone convergence yields $E_{Q}(A_T)=E_P(Z_{-} \mal A_T)$ as claimed. \ep

\end{appendix}

\end{document}